\documentclass{amsart}[12]

\usepackage{amssymb}
\usepackage{graphicx}
\usepackage[english]{babel}
\usepackage[pdftitle={},
  pdfauthor={D. Arjmand},
  pdffitwindow=true,
  colorlinks=true,
  urlcolor=blue,
  linkcolor=red,
  citecolor=red,
  anchorcolor=red]{hyperref}
\usepackage[numbers]{natbib}
\usepackage{enumerate}
\usepackage{amsaddr}
\usepackage{marginnote}
\usepackage{ulem}

\numberwithin{equation}{section}
\numberwithin{table}{section}
\numberwithin{figure}{section}

\newcommand{\e}{\varepsilon}
\newcommand{\vectorn}[1]{\ensuremath{ \boldsymbol{#1} }}
\newcommand{\tensor}[1]{\ensuremath{ \boldsymbol{#1} }}
\newcommand{\dif}{\ensuremath{ \mathrm{d} }}

\newtheorem{Remark}{Remark}

\begin{document}

\title[Multiscale modelling of long-range interactions]{Modelling long-range interactions in multiscale simulations of ferromagnetic materials}

\author[D. Arjmand]{Doghonay Arjmand}
\address{ANMC, Section de Math\'{e}matiques, \'{E}cole polytechnique f\'{e}d\'{e}rale de Lausanne, CH-1015 Lausanne, Switzerland}
\email{doghonay.arjmand@epfl.ch}
\author[M. Poluektov]{Mikhail Poluektov}
\address{International Institute for Nanocomposites Manufacturing, WMG, University of Warwick, Coventry CV4 7AL, UK} 
\email{m.poluektov@warwick.ac.uk}
\author[G. Kreiss]{Gunilla Kreiss}
\address{Division of Scientific Computing, Department of Information Technology, Uppsala University, SE-751 05 Uppsala, Sweden}
\email{gunilla.kreiss@it.uu.se}




\begin{abstract}
Atomistic-continuum multiscale modelling is becoming an increasingly popular tool for simulating the behaviour of materials due to its computational efficiency and reliable accuracy. In the case of ferromagnetic materials, the atomistic approach handles the dynamics of spin magnetic moments of individual atoms, while the continuum approximations operate with volume-averaged quantities, such as magnetisation. One of the challenges for multiscale models in relation to physics of ferromagnets is the existence of the long-range dipole-dipole interactions between spins. The aim of the present paper is to demonstrate a way of including these interactions into existing atomistic-continuum coupling methods based on the partitioned-domain and the upscaling strategies. This is achieved by modelling the demagnetising field exclusively at the continuum level and coupling it to both scales. Such an approach relies on the atomistic expression for the magnetisation field converging to the continuum expression when the interatomic spacing approaches zero, which is demonstrated in this paper.
\end{abstract}

\maketitle

\section{Introduction}

There are multiple ways of describing the physics of magnetic materials. At the smallest scale, the spin and orbital movements of electrons are modelled by electronic structure calculations. At a larger scale, the rapid subatomic variations are averaged out and the interaction of spin magnetic moments of individual atoms is simulated, often by using parametrised interactions obtained from a smaller scale. The interaction of atomic spins is described by a system of coupled nonlinear ordinary differential equations (ODEs). At the macroscopic scale, nonlinear partial differential equations (PDEs) are used to describe the evolution of volume-averaged quantities. The choice of a computational approach depends not only on the scale of application, but also on the required computational efficiency. The atomistic models, although relatively accurate, are prohibitively expensive to solve, whereas continuum models are computationally efficient but may lack certain accuracy.

In contrast to targeting a single scale, multiscale modelling strategies potentially offer the accuracy of atomistic models combined with the efficiency of the macroscopic models. All multiscale models can be categorised into sequential (one-way coupling) and concurrent (two-way coupling) methods. The concurrent multiscale models, in turn, can be separated into two groups --- the partitioned-domain and the hierarchical approaches \cite{Tadmor2011}. The latter is referred to as upscaling approach in this paper.

In the partitioned-domain approaches, the entire physical domain is split in regions represented by the atomistic and the continuum models with an explicit interface between them. The information exchange takes place at the interface and the major challenges for these models are handling non-local atomistic interactions and averaging fast atomistic variations at the interface. In the upscaling approach \cite{Abdulle_E_Engquist2012}, atomistic models are solved locally to find unknown macroscopic quantities in an initially incomplete macroscopic model. The macro model is then evolved over the entire computational domain. The upscaling strategy, which is considered in this paper, follows the general framework of the heterogeneous multiscale method (HMM) and uses a two-way coupling between the atomistic and macro models, where the atomistic simulations use the macro data as initial or boundary conditions, while the macro model uses the information coming from local computations of the atomistic model. 

The domain partitioning approach is intended for cases when an interaction of a magnetic structure, e.g. a domain wall or a skyrmion, with an isolated heterogeneity, e.g. a crystallographic defect, is studied. In this case, the region of interest in modelled using the atomistic approach, while in the rest of the computational region, the continuum model is used. Such approach relies on the continuum model to be well-defined, i.e. derivable from the atomistic model up to a small course-graining error, which might be neglected far from the region of interest. This is the case at relatively small temperatures and for homogeneous materials. The upscaling approach, on the other hand, is intended for cases when the material microstructure is heterogeneous, but representable using periodically stacked representative volume elements (RVEs). In this case, the continuum model is not well-defined and must be obtained by upscaling. The upscaling strategy is also applicable in other cases, when the continuum model is not well-defined, e.g. magnetic structures under a high temperature or a high-frequency external fields. In what follows, a short overview of applications of the partitioned-domain and upscaling approaches in relation with multiscale problems for ferromagnetic materials is given. 

Construction of multiscale models for magnetic materials is a rapidly developing field and a number of partitioned-domain techniques have been proposed in the past \cite{GarciaSanchez2005,Jourdan2008,Andreas2014,DeLucia2016,Poluektov2016,Poluektov2018}. For the overview and the comparison of various methods the reader is referred to exhaustive review articles \cite{Miller2009,Luskin2013} discussing partitioned-domain methods in general and \cite{Hertel2018} discussing application of multiscale models to magnetic materials in particular. As mentioned above, a major challenge for the partitioned-domain approach is constructing an interface without introducing surplus artefacts into simulations. In \cite{Poluektov2016}, the problem of high-frequency wave reflections from the atomistic-continuum interface has been addressed by introducing additional numerical damping, while in \cite{Poluektov2018}, a way of handling non-local interatomic interactions at the interface, by introducing a transition zone with partially coarse-grained interactions, has been suggested.

In terms of upscaling approaches, a way of constructing a macroscopic model of ferromagnetic materials, which is fully coupled to an atomistic model, has recently been reported in \cite{Arjmand_etal_2016}, where an analysis of the dynamics of a single particle and a chain of particles subjected to a high-frequency external field was given. The extension of the method to problems at elevated temperatures was addressed in \cite{APK_Temp_2017}. In the case of a non-zero temperature, the macroscopic magnetisation vector field, which is the volume average of atomistic spin magnetic moments, has temperature-dependent length. In \cite{APK_Temp_2017}, it has been shown that the upscaling method accurately captures the reduced magnetisation length at the macroscopic scale.

In the case of modelling magnetic materials, there is an additional challenge for multiscale models --- the existence of long-range dipole-dipole interactions between atomic spins \cite{Aharoni1996}. These interactions cannot be handled in the same way as the short-range interatomic interactions, since this would require unreasonably large padding and/or transition zones in the partitioned-domain approach and unreasonably large microscopic domains in the upscaling approach. Such treatment of long-range interactions diminishes all advantages of multiscale approaches. Therefore, these interactions should be handled in a conceptually different way --- using a continuum approach. The aim of this paper is to present an efficient strategy to include the long-range interactions in the partitioned-domain strategy based on \cite{Poluektov2016,Poluektov2018} and the upscaling formalism developed in \cite{Arjmand_etal_2016,APK_Temp_2017}.

This paper is organised as follows. In Section \ref{Sec_MathModels}, mathematical models governing the behaviour at the atomistic and the continuum scales are described and a convergence study is carried out to quantify the approximation errors in relation to modelling the long-range interactions. In Section \ref{Sec_MultiscaleModel}, the multiscale models based on the partitioned-domain and the HMM frameworks, are presented. Finally, numerical results are provided in Section \ref{Sec_NumericalResults} to demonstrate the accuracy of the proposed methods. 

\section{Mathematical models at different scales}
\label{Sec_MathModels}

\subsection{Atomistic spin dynamics}

At the atomistic scale, the mathematical model is the atomistic Landau-Lifshitz-Gilbert equation \cite{Bergqvist2013,Evans2014}, which is given by 
\begin{align}
  &\frac{\dif}{\dif t} \vectorn{m}_i = -\beta_\mathrm{L} \vectorn{m}_i \times \vectorn{H}_i - \alpha_\mathrm{L} \vectorn{m}_i
    \times \left( \vectorn{m}_i \times \vectorn{H}_i \right) , \quad \left|\vectorn{m}_i\right| = 1,
  \label{eq:ASD_LLG} \\
  &\beta_\mathrm{L} = \frac{\gamma}{1+\lambda^2} , \quad \alpha_\mathrm{L} = \frac{\gamma\lambda}{1+\lambda^2} ,
  \label{eq:ASD_param} \\
  &\vectorn{H}_i = \frac{1}{\mu}\left( \sum_j J_{ij} \vectorn{m}_j \right) +
    \frac{1}{\mu} \tensor{K}_\mathrm{a} \cdot \vectorn{m}_i +
    \vectorn{H}_\mathrm{e} + \vectorn{h}_i  + \vectorn{H}_{\mathrm{a},i},
  \label{eq:ASD_H}
\end{align}
where $\gamma$ is the gyromagnetic ratio, $\lambda$ is the phenomenological damping constant, $\vectorn{m}_i$ is the direction of spin magnetic moment, $\mu$ is the length of spin magnetic moment, $J_{ij}$ are constants of Heisenberg exchange interaction between atoms $i$ and $j$, $\tensor{K}_\mathrm{a}$ is the anisotropy tensor and $\vectorn{H}_\mathrm{e}$ is the external field. Thermal excitations are taken into account by adding a stationary stochastic field with the following statistical properties: 
\begin{align}
  &\left\langle h_{i\rho}\left(t\right) \right\rangle = 0 , \quad
    \left\langle h_{i\rho}\left(t\right) h_{j\nu}\left(s\right) \right\rangle = 2 D \delta_{ij} \delta_{\rho\nu} \delta\left(t-s\right) ,
  \label{eq:ASD_noise} \\
  &D = k_\mathrm{B}T \frac{\lambda}{\mu\gamma} ,
\end{align}
where $\rho$ and $\nu$ are the Cartesian coordinates of $\vectorn{h}_i$, $k_\mathrm{B}$ is the Boltzmann constant and $T$ is temperature. Finally, the term $\vectorn{H}_{\mathrm{a},i}$ is a demagnetising field, which originates from long-range dipole-dipole interactions between spin magnetic moments, and is given by \cite{Aharoni1996} 
\begin{equation}
  \vectorn{H}_{\mathrm{a},i} = \mu_0 \mu \left( -\dfrac{\vectorn{m}_i}{3 V_\mathrm{a}} + \dfrac{1}{4 \pi} \sum_{r_{ij} \neq 0} 
    \left( \dfrac{3 \vectorn{m}_j \cdot \vectorn{r}_{ij} \vectorn{r}_{ij}}{r_{ij}^{5}} -
    \dfrac{\vectorn{m}_j}{r_{ij}^3} \right) \right) , \quad
    r_{ij} = \left| \vectorn{r}_{ij} \right| ,
  \label{eq:LongRangeAtomistic}
\end{equation}
where $\vectorn{r}_{ij}$ is the vector connecting atoms $i$ and $j$, and $V_{\mathrm{a}}$ is the volume occupied by a single atom\footnote{For crystallographic lattice with cubic stacking, $V_{\mathrm{a}} = a^3$, where $a$ is the distance between two neighbouring atoms.}. Parameters $\lambda$, $\mu$, $J_{ij}$ and $\tensor{K}_\mathrm{a}$ can be computed from electronic structure calculations \cite{Eriksson2016} and are considered to be constant for a given material, and $\mu_0$ is the permeability of free space.  

\subsection{Continuum models for magnetisation dynamics}
\label{sec:cont}

In this section, two different continuum models are presented. In the first part, a well-known continuum model from the classical micromagnetic theory is provided. It is followed by an alternative continuum model based on the upscaling approach.

\subsubsection{Continuum model from classical micromagnetism}

At the continuum scale, the magnetisation dynamics is modelled by the following nonlinear partial differential equation  \cite{Aharoni1996,Cimrak2008}:
\begin{align}
  &\frac{\partial}{\partial t} \vectorn{M}\left(t,\vectorn{x}\right) = -\beta_\mathrm{L} \vectorn{M} \times \vectorn{H} -
    \alpha_\mathrm{L} \vectorn{M} \times \left( \vectorn{M} \times \vectorn{H} \right) , \quad \left|\vectorn{M}\right| = 1 ,
  \label{eq:C_LLG} \\
  &\vectorn{H}\left(t,\vectorn{x},\vectorn{M}\right) = \frac{1}{\mu} \tensor{A}_\mathrm{e} : \nabla\nabla \vectorn{M} +
    \frac{1}{\mu} \tensor{K}_\mathrm{a} \cdot \vectorn{M} + \vectorn{H}_\mathrm{e} + \vectorn{H}_\mathrm{c},
  \label{eq:C_H}
\end{align}
where $\vectorn{M}$ is the normalised magnetisation field and $\beta_\mathrm{L}$ and $\alpha_\mathrm{L}$ are the same coefficients as used in \eqref{eq:ASD_LLG}. At zero temperature, exchange tensorial\footnote{Here, the standard tensor notation is used, where the tensor product of two vectors is denoted as $\vectorn{a}\vectorn{b}$, which results in a second-order tensor. The double inner product of two second-order tensors is denoted as $\tensor{A}:\tensor{B} = \sum_i \sum_j A_{ij} B_{ji}$, which results in a scalar.} parameter $\tensor{A}_\mathrm{e}$ can be obtained directly from the atomistic parameters:
\begin{equation} 
  \tensor{A}_\mathrm{e} = \frac{1}{2} \sum_{j \neq i} J_{ij} \vectorn{r}_{ij} \vectorn{r}_{ij} ,
  \label{eq:C_exch_A}
\end{equation}
where $\vectorn{r}_{ij}$ is the vector connecting atoms $i$ and $j$. The sum is evaluated over all atoms with which atom $i$ interacts. In \eqref{eq:C_exch_A}, tensor $\tensor{A}_\mathrm{e}$ is assumed to be spatially constant. Since the anisotropy term is local, the same anisotropy tensor $\tensor{K}_\mathrm{a}$ is used in the continuum and the atomistic equations.

The demagnetising field is denoted as a vector field $\vectorn{H}_{\mathrm{c}}(\vectorn{x})$ for all $\vectorn{x} \in \Omega_{\mathrm{in}}$, where $\Omega_{\mathrm{in}} \subset \mathbb{R}^{3}$ defines the interior region of a given magnetic material. The outer region is defined as $\Omega_{\mathrm{out}} = \mathbb{R}^{3}/\Omega_{\mathrm{in}}$. This field is given by
\begin{align}
  &\vectorn{H}_{\mathrm{c}} = -c_\mathrm{L} \nabla U , \quad\quad
    c_\mathrm{L} = \frac{\mu_0 \mu}{V_{\mathrm{a}}} ,
  \label{eq:C_HLR} \\
  &U\left(\vectorn{x}\right) := \begin{cases} U_\mathrm{in}\left(\vectorn{x}\right) ,
    & \vectorn{x} \in \Omega_\mathrm{in} , \\
    U_\mathrm{out}\left(\vectorn{x}\right) ,
    & \vectorn{x} \in \Omega_\mathrm{out} ,
    \end{cases} 
\end{align}
where $U$ is the solution of the following PDE \cite{Aharoni1996}:
\begin{equation}
  \left\{
  \begin{array}{lll}
  \Delta U_{\mathrm{in}}\left(\vectorn{x}\right) = \nabla \cdot \vectorn{M}\left(\vectorn{x}\right), &
    \text{in } \Omega_{\mathrm{in}} , \\
  \Delta U_{\mathrm{out}}\left(\vectorn{x}\right) = 0, &
    \text{in } \Omega_{\mathrm{out}} , \\
  U_{\mathrm{in}} = U_{\mathrm{out}}, &
    \text{on } \partial\Omega_{\mathrm{in}} , \\
  \partial_n U_{\mathrm{in}} - \partial_n U_{\mathrm{out}} =  \vectorn{M} \cdot \vectorn{n}, &
    \text{on } \partial\Omega_{\mathrm{in}} , \\
  \left| r U_{\mathrm{out}} \right| \to 0 , \quad \left| r^2 \nabla U_{\mathrm{out}} \right| \to 0 , &
    \text{at } r \to \infty , &
    \text{where } r = \left| \vectorn{x} \right| .
  \end{array}
  \right.
  \label{eq:EquationsForPotential}
\end{equation}
Here, the coefficient $c_\mathrm{L}$ is introduced to preserve the physical meaning of the magnetisation as the `density' of spin magnetic moments, while the continuum equation is based on the normalised magnetisation, i.e. $\left|\vectorn{M}\right| = 1$.

The continuum magnetisation field $\vectorn{M}$ is equal to the normalised ensemble average of the volume average of the atomic spin magnetic moments $\vectorn{m}_i$. The normalisation is introduced due to the nature of the LLG equation, as the unit-length vectors are usually used in the formulations. In general, such atomistic-continuum transition introduces an error to the solution, which is dependent on interatomic spacing and on the magnetisation gradient \cite{Poluektov2018}.

It must be noted that at finite temperatures, the continuum model must be modified. These modifications differ depending on the approach and are discussed in \cite{APK_Temp_2017}. However, it has been shown that even with the modifications, the continuum model cannot approach the atomistic model with a predefined accuracy at finite temperatures, i.e. there is always a finite temperature-dependent error. This is one of the reasons for introducing an alternative continuum model based on upscaling.

\subsubsection{A continuum model based on upscaling}

The basic idea behind upscaling approaches is to start by assuming a macro model, in which certain quantities are unknown and must be obtained from a given microscopic model. The form of the macro model usually requires some knowledge about the physical laws that govern the evolution of macroscopic variables. A macro model in the form of
\begin{equation}
  \dfrac{\partial}{\partial t} \vectorn{M} = -\vectorn{F}\left(t, \vectorn{x} ,\vectorn{M}\right) -
    \dfrac{\alpha_{\mathrm{L}}}{\beta_{\mathrm{L}}} \vectorn{M} \times \vectorn{F}\left(t, \vectorn{x}, \vectorn{M}\right) 
  \label{Eqn_HMM_Macro_Model}
\end{equation}
has been proposed and analysed in \cite{Arjmand_etal_2016}. In this macro model, the term $\vectorn{F}$ is an unknown quantity, which is then upscaled using the local microscopic equation. On the other hand, the macroscopic response, $\vectorn{M}$, is obtained by using a suitable time discretisation. While designing such upscaling strategies, one important issue is the synchronisation of the micro problems using the coarse-scale variables, which is achieved by assigning suitable initial and boundary conditions for the micro problems. Note that the long-range field does not appear explicitly in the macro model \eqref{Eqn_HMM_Macro_Model}, but the macro model should capture the effect of the long-range interactions. Later, in Section \ref{subsec:HMM}, it is shown that this can be achieved by including the long-range field in the microscopic models without computing the computationally expensive atomistic dipole-dipole interactions.

\subsection{Quantification of the errors associated with approximations of the demagnetising field}

It is well-known that from the purely mathematical point of view, the atomistic demagnetisation field \eqref{eq:LongRangeAtomistic} is consistent with the continuum demagnetisation field \eqref{eq:C_HLR}. This can be seen from the representation of the continuum PDE for $U$ using the Green's function \cite{Aharoni1996}, subsequent application of the divergence theorem and substitution into expression for $H_\mathrm{c}$, which gives
\begin{equation*}
  \vectorn{H}_\mathrm{c}\left( \vectorn{r} \right) = \frac{c_\mathrm{L}}{4\pi} \int_{\Omega_\mathrm{in}} \left(
    \frac{3 \vectorn{M} \left( \vectorn{r}' \right) \cdot
    \left( \vectorn{r} - \vectorn{r}' \right)\left( \vectorn{r} - \vectorn{r}' \right)}{\left| \vectorn{r} - \vectorn{r}' \right|^5} -
    \frac{\vectorn{M}\left( \vectorn{r}' \right)}{\left| \vectorn{r} - \vectorn{r}' \right|^3} \right) \dif V' .
\end{equation*}
The atomistic expression \eqref{eq:LongRangeAtomistic} can be seen as a discretisation of the integral expression for $\vectorn{H}_\mathrm{c}$. 

Since construction of the multiscale models with the long-range interaction relies on the consistency between the atomistic and the continuum expressions for the demagnetisation field, the aim of this section is to demonstrate this consistency numerically and to investigate the convergence rates of the errors related to a) the approximation error, quantifying the error between the continuum field $\vectorn{H}_{\mathrm{c}}$ and the atomistic field $\vectorn{H}_{\mathrm{a},i}$, and b) the geometric error, which is due to neglecting the far-field particles in the computation of $\vectorn{H}_{\mathrm{a},i}$. For the sake of comparison, a brief derivation of a 1D solution for the magnetic potential is provided here.

\subsubsection{Analytical solution for the magnetic potential in 1D}
\label{Sec_OneDimAnal}

Consider the domain $\Omega_{\mathrm{in}} = [0,R]\times\mathbb{R}^2$, which is bounded in the $x_1$ direction and infinite in $x_2$ and $x_3$ directions. Assume also a magnetisation vector field $\vectorn{M} = \vectorn{M}(x_1)$, which is a function of the first coordinate only. Then it follows that 
\begin{equation}
  \Delta U_{\mathrm{in}}(\vectorn{x}) = \partial_{x_1} M_1(x_1) , \quad
  \Delta U_{\mathrm{out}}(\vectorn{x}) = 0 .
\end{equation}
This is a one-dimensional problem, i.e. $U_\mathrm{in}$ and $U_\mathrm{out}$ depend only on $x_1$, since the magnetisation $\vectorn{M}$ depends only on $x_1$. Hence the Laplace operator $\Delta $ reduces to the second derivative in $x_1$, and the following solution is directly obtained:
\begin{align}
  &U_{\mathrm{in}}(x_1) = \int_{0}^{x_1} M_1(s_1) \; \dif s_1 + c_1 x_1 + c_0 , 
  \label{Eqn_Uin} \\
  &U_{\mathrm{out}}(x_1) = b_0 + b_1 x_1 .
\end{align}
For $U_{\mathrm{out}}$ to be bounded as $x_1 \to \infty$, it is necessary that $\partial_{x_1} U_{\mathrm{out}}(x_1)  = b_1 = 0$. This implies that $U_{\mathrm{out}}$ is a constant. Moreover, since there is a jump in the derivative of $U$ at interfaces $x_1 = 0$ and $x_1 = R$, one obtains 
\begin{align*}
  &\partial_{x_1} U_{\mathrm{in}}(R) = \partial_{x_1} U_{\mathrm{out}}(R) + M_1(R) = M_1(R) , \\
  &\partial_{x_1} U_{\mathrm{in}}(0) = \partial_{x_1} U_{\mathrm{out}}(0) + M_1(0) = M_1(0) .
\end{align*}
Using these equalities in the equation \eqref{Eqn_Uin} yields $c_1 = 0$. Hence, $U_{\mathrm{in}}$ becomes
\begin{equation*}
  U_{\mathrm{in}}(x_1) = \int_{0}^{x_1} M_1(s_1) \; \dif s_1 + c_0 .
\end{equation*}   
This leads to
\begin{equation}
  \vectorn{H}_{\mathrm{c}}(\vectorn{x}) = -c_\mathrm{L}\nabla U_{\mathrm{in}}(\vectorn{x}) = -c_\mathrm{L} M_1(x_1) \vectorn{e}_1 , \quad \forall \vectorn{x} \in \Omega_{\mathrm{in}} .
  \label{eq:HL_Formula}
\end{equation}

It must be noted that in the derivation above, $U_{\mathrm{out}}$ does not necessarily decay to zero, but remains bounded for all $x_1$. However, the derivative $\partial_{x_1} U_{\mathrm{out}}$ must decay to zero at infinity for a well-defined solution. This is related to the one-dimensional nature of the problem. Namely, the Green's function in 1D does not decay to zero. In a truly three dimensional problem, the solution itself also has to decay at infinity. 

Analytical solutions are also available in a few specific three-dimensional domains with a uniform magnetisation over the domain \cite{Aharoni1996}.

\subsubsection{A qualitative comparison between $\vectorn{H}_{\mathrm{c}}$ and $\vectorn{H}_{\mathrm{a},i}$}

In what follows, the aim is to demonstrate that the terms $\vectorn{H}_{\mathrm{a},i}$ and $\vectorn{H}_{\mathrm{c}}$, which are given by \eqref{eq:LongRangeAtomistic} and \eqref{eq:C_HLR}, respectively, match qualitatively. The domain $\Omega_{\mathrm{in}} = [0,1]\times\mathbb{R}^2$ is taken and magnetisation $\vectorn{M} = \vectorn{M}(x_1)$ is assumed to be dependent on the first coordinate only. From \eqref{eq:HL_Formula}, the continuum solution is given by $\vectorn{H}_{\mathrm{c}}(\vectorn{x}) = -c_\mathrm{L} M_1(x_1) \vectorn{e}_1$. For the purpose of this example, $c_\mathrm{L} = 1$ is assumed. To compute the atomistic field $\vectorn{H}_{\mathrm{a},i}$ via equation \eqref{eq:LongRangeAtomistic}, first, a truncation of the domain $\Omega_{\mathrm{in}}$ is required. Therefore, the following hyper-rectangle is defined:
\begin{equation*}
  \Omega_{R} = [0,1]\times[-R/2,R/2]^2 .
\end{equation*}
Next, the atomistic lattice is defined as a uniform discretisation of $\Omega_{R}$, where the step size of the discretisation is chosen to be equal to the interatomic distance $a$. Namely,
\begin{equation}
  \Omega_{R,a} := \left\{\left( x_1^{i} = i a ,\quad x_2^{j} = -R/2 + j a,\quad x_3^{k} = -R/2 + k a \right) \right\} ,
  \label{eq:AtomisticLattice}
\end{equation}
where 
\begin{equation*}
  i = 0, \ldots, N_1, \quad N_1 a = 1, \quad j,k = 0, \ldots, N_2, \quad N_2 a = R .
\end{equation*}

The atomistic field is computed over the lattice \eqref{eq:AtomisticLattice}, which contains $(N_1+1) (N_2+1)^2$ number of atoms. It is also assumed that the magnetic moments are given by 
\begin{equation*}
  \vectorn{m}_i = \frac{1}{\sqrt{2}} \left( \cos(2 \pi x_1^{i}) \vectorn{e}_1 + \sin(2 \pi x_1^{i}) \vectorn{e}_2 + \vectorn{e}_3 
    \right) , \quad \forall i .
\end{equation*}

\begin{figure}
  \begin{center}
    \includegraphics[scale=0.4]{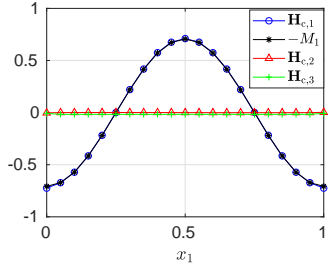}
  \end{center}
  \caption{A comparison of the atomistic field $\vectorn{H}_{\mathrm{a}}$ and the continuum field $\vectorn{H}_{\mathrm{c}}$ for $a=0.05$ and $R=20$. The $x_1,x_2,x_3$ components of the atomistic field are plotted against the $x_1$ component of the continuum field, which is $-M_1$.}
  \label{fig:Comp_Atom_Cont}
\end{figure}

Figure \ref{fig:Comp_Atom_Cont} confirms a qualitative match between the atomistic expression \eqref{eq:LongRangeAtomistic} for $\vectorn{H}_{\mathrm{a,i}}$ and the continuum equation \eqref{eq:C_HLR} for $\vectorn{H}_{\mathrm{c}}$. It is evident, from equation \eqref{eq:LongRangeAtomistic}, that the computed atomistic field $\vectorn{H}_{\mathrm{a},i}$ depends on the truncation length $R$, as well as the choice of the atomic distance $a$. This dependency results in a small difference between the atomistic and the continuum fields depicted in Figure \ref{fig:Comp_Atom_Cont}. The next subsection focuses on studying convergence rates for the errors with respect to $a$ (the approximation error) and $R$ (the geometric error). Note that the $x_2$ and $x_3$ components of the continuum field are equal to zero and hence are not included in the figure. 

\subsubsection{The approximation error}
\label{subsec:ApproximationError}

Given a magnetic body $\Omega \subset \mathbb{R}^3$ filled with a number of atoms, the continuum field $\vectorn{H}_{\mathrm{c}}(\vectorn{x}_i)$, $\vectorn{x}_i \in \Omega$ is obtained as the limit of the atomistic field $\vectorn{H}_{\mathrm{a},i}$ when $a \to 0$, i.e.
\begin{equation*}
  \lim_{a \to 0} \vectorn{H}_{\mathrm{a},i} = \vectorn{H}_{\mathrm{c}}(\vectorn{x}_i) .
\end{equation*}
The aim here is to analyse the rate of convergence by considering the volume $[0,1]^2 \times [-1,1]$ in $\mathbb{R}^3$  and computing the atomistic field $\vectorn{H}_{\mathrm{a},i}$ for decreasing values of $a$. The magnetic moments are assumed to be uniform everywhere and pointing in the $x_1$ direction,
\begin{equation*}
  \vectorn{m}_i = \vectorn{e}_1 .
\end{equation*}
In particular, a sequence of atomic distances given by $a_k = 2^{k} a_{\mathrm{min}}$, where $k=0,1,\ldots,5$ and $a_{\mathrm{min}} = 0.00625$ is used, and the following differences are recorded.
\begin{equation*}
  E_{k} := \left| \vectorn{H}_{\mathrm{a},i}^{a_{k+1}} - \vectorn{H}_{\mathrm{a},i}^{a_{k}} \right|,
    \quad k=0,1,\ldots,4.
\end{equation*}
The convergence rate can then be obtained by computing $\{ s_k \}_{k=1}^{4}$, where
\begin{align*}2^{s_k} = \dfrac{E_k}{E_{k-1}}.
\end{align*}
The values of $E_k$ and $s_k$ are shown in Table \ref{Table_Approximation_Error}, and a first order convergence rate with respect to the interatomic spacing for the approximation error is observed for this specific example.
 
\begin{table}
\begin{tabular}{l | l | l |l |l |l}
 $k$     & $4$ & $3$ & $2$  &  $1$ & $0$ \\ \hline
 $E_{k}$ & $ 0.1605 $ & $ 0.1019 $ & $ 0.0619 $  & $ 0.0324 $ & $ 0.0165 $  \\ \hline
 $s_k$   & $ 0.6555 $ & $ 0.7173 $ & $ 0.9344 $  & $ 0.9713 $ & $ $ 
\end{tabular}
\bigskip
\caption{The approximation error. Decrease of $k$ corresponds to the decrease of the interatomic spacing, i.e. for $k=4$, $a=0.1$, while for $k=0$, $a=0.00625$. Here, $s_k$ is the numerical approximation of the convergence rate with respect to the interatomic spacing.}
\label{Table_Approximation_Error}
\end{table}

\subsubsection{The geometric error}

When computing the atomistic field $\vectorn{H}_{\mathrm{a},i}$, one often has to deal with large computational geometries relative to the atomic distance $a$. When the size of the magnetic body is large, computation of $\vectorn{H}_{\mathrm{a},i}$ via the summation formula \eqref{eq:LongRangeAtomistic} is unreasonably expensive. One strategy can be ignoring the atoms, which are located far from atom $i$, implying a truncation of the computational geometry. In this subsection, the goal is to understand the decay of the error, which arises from truncating the computational geometry. For this, the domain $[0,1]^2 \times [-R,R]$ is uniformly discretised using the interatomic distance $a = 0.1$. For the computations, it is assumed that 
\begin{equation*}
  \vectorn{m}_i = \frac{1}{\sqrt{2}} \left( \cos(2 \pi x_1^{i}) \vectorn{e}_1 + \sin(2 \pi x_1^{i}) \vectorn{e}_2 +
    \vectorn{e}_3 \right) , \quad \forall i .
\end{equation*}
The convergence is studied for $R_k = 2^{k}, k=0,1,\ldots,4$, and the errors 
\begin{equation*}
  E_{k} := \left| \vectorn{H}_{\mathrm{a},i}^{R_{k+1}} - \vectorn{H}_{\mathrm{a},i}^{R_{k}} \right|,
    \quad k=0,1,2,3
\end{equation*}
are recorded. To find the rate $s$ of the geometric error, which is assumed to be $O(R^{-s})$, the following ratios are computed.
\begin{equation*}
2^{-s_k} = \dfrac{E_k}{E_{k-1}}.
\end{equation*}
The results summarised in Table \ref{Table_Geometric_Error}, show a second order convergence rate for the geometric error. 

\begin{table}
\begin{tabular}{l | l | l |l |l }
 $k$     & $0$ & $1$ & $2$  &  $3$  \\ \hline
 $E_{k}$ & $ 0.5992 $ & $ 0.1963 $ & $ 0.0538 $  & $ 0.0138 $   \\ \hline
 $s_k$   & $ 1.6096 $ & $ 1.8663 $ & $ 1.9545 $  & $ $  
\end{tabular}
\bigskip
\caption {The geometric error. Increase of $k$ corresponds to the increase of the truncation radius, i.e. for $k=0$, $R=1$, while for $k=3$, $R=8$. Here, $s_k$ is the numerical approximation of the convergence rate with respect to the cutoff radius.}
\label{Table_Geometric_Error}
\end{table}

\section{Multiscale modelling}
\label{Sec_MultiscaleModel}

\subsection{Partitioned-domain approach}

\subsubsection{Energy-based and force-based coupling}

\begin{figure}
  \begin{center}
    \includegraphics[scale=0.8]{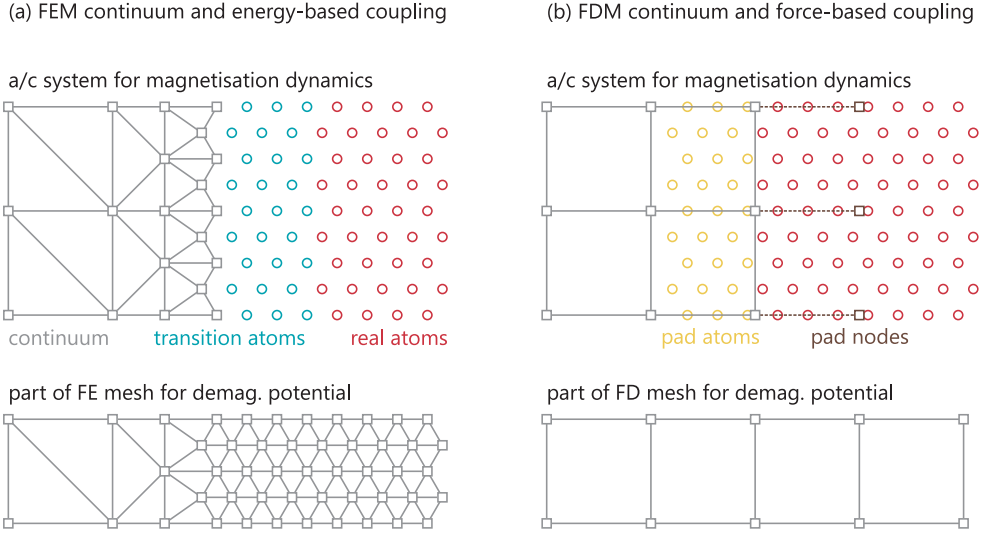}
  \end{center}
  \caption{Schematic representation of different versions of partitioned-domain atomistic-continuum (a/c) coupling. The a/c coupling is only used for solving the magnetisation dynamics, while the demagnetisation field is solved using exclusively the continuum approach, with the continuum mesh extended to the entire computational domain, covering the atomistic region, i.e. upper a/c systems and lower meshes are discretisations of the same physical domain. The demagnetisation field is used in the LLG equations, while magnetisation is used when solving for the magnetic potential.}
  \label{fig:scheme}
\end{figure}

In the multiscale partitioned-domain coupling approach considered in this paper, the entire computational region is split into two subregions --- the atomistic and the continuum domains. There is a ``sharp'' atomistic-continuum interface between these two regions, as illustrated in figure \ref{fig:scheme}.

It is well-known that all partitioned-domain methods can be separated into two conceptually distinct groups --- the energy-based coupling and the force-based coupling \cite{Tadmor2011}. In the energy-based coupling, the total energy functional of the system is written and forces or torques are derived from it, in the case of modelling deformation or magnetism, respectively. The continuum and the atomistic equations are discretised in time and are advanced together as a single unified system. Thus, in the absence of damping and when an energy-conserving time-stepping method is used, the total energy of the system is conserved. In the force-based coupling, the continuum and the atomistic regions are advanced separately, while exchanging the boundary conditions via the padding atoms and the interface. This ensures that the correct solution is transferred between the regions; however, the total energy of the system is not well-defined in this case.

The continuum region can be discretised using either the finite-element method (FEM) or the finite-difference method (FDM). FEM has an advantage of creating an interface that is conforming to the atomistic lattice, which allows constructing an energy-conserving coupling method \cite{Poluektov2018}; however, it has a more complex implementation than FDM. In the case of FDM-discretised continuum, the coupling requires boundary conditions for the continuum region, which can be obtained from additionally constructed pad nodes, where the solution is obtained by volume-averaging of the atomistic solution.

Most energy-based methods have a disadvantage of having numerical artefacts at the atomistic-continuum interface, which are referred to as ``ghost-forces'' in the case of modelling deformation or ``ghost-torques'' in the case of magnetism, which emerge due to a non-local interatomic interaction. These artefacts can only be removed by employing complex methods of constructing transition zones at the atomistic-continuum interface \cite{Ortner2014,Poluektov2018}. Thus, the energy-based coupling requires modification of atoms close to the interface into transition atoms to remove ghost-forces/torques. The force-based coupling, on the other hand, requires construction of pad atoms to provide the boundary conditions for the atomistic region, which is simpler both computationally and in terms of implementation. The disadvantage of the force-based methods is the absence of the well-defined total energy of the system. Thus, the energy-based and the force-based methods have a somewhat different scope. For the discussion of the energy-based vs the force-based methods, the reader is referred to \cite{Tadmor2011}.

\subsubsection{Including the long-range interactions}

In the case of magnetic materials, the long-range dipole-dipole interactions are by definition non-local. However, they cannot be handled in the same way as, for example, the exchange interactions, which can be truncated at a relatively short range. In the energy-based approach, the transition zone that removes the ghost-torques must be of the same width as the interaction distance \cite{Poluektov2018}, which makes it impractical to create such a zone. In the force-based approach, the width of the region with padding atoms must also be larger than the interaction distance, which also renders it impractical. Therefore, the only solution is to use a conceptually different handling of these interactions.

The idea of handling long-range dipole-dipole interactions in multiscale models of magnetic materials is to model these interactions using an exclusively continuum approach. To the best knowledge of the authors, this idea has not been suggested so far in the context of the multiscale approach in application to the magnetisation dynamics. Since the atomistic and the continuum domains occupy different spatial domains, an auxiliary computational mesh that covers the entire physical region must be introduced, as illustrated in figure \ref{fig:scheme}. The continuum equation for the magnetic potential \eqref{eq:EquationsForPotential} is then solved on this auxiliary mesh. Within the region of this auxiliary mesh that covers the atomistic region, the atomistic demagnetising field $\vectorn{H}_{\mathrm{a},i}$ is equated to the continuum demagnetising field $\vectorn{H}_\mathrm{c}$, while the magnetisation $\vectorn{M}$ in \eqref{eq:EquationsForPotential} is, in turn, equated to the atomistic solution $\vectorn{m}_i$, which might require interpolation and/or volume-averaging that is discussed below. This idea relies on the convergence of the atomistic expression for the demagnetising field \eqref{eq:LongRangeAtomistic} to the continuum expression \eqref{eq:C_HLR} as $a \to 0$.

There will be differences, however, depending on whether the approach is energy-based or force-based and whether FEM or FDM is used for the continuum region. In the energy-based approach and FEM-discretised continuum with the finite-element mesh refined down to the atomistic lattice, figure \ref{fig:scheme}a, an auxiliary mesh can be constructed, which contains nodes that exactly coincide with the continuum nodes within the continuum region and that exactly coincide with the atomistic positions within the atomistic region. This ensures that interpolation of the solution and the demagnetising field is avoided. Thus, when an energy-conserving time-stepping method is used, such a system will be energy-conserving.

In the force-based approach and FDM-discretised continuum with the structured mesh, figure \ref{fig:scheme}b, the auxiliary mesh will be also structured and will be an extension of the continuum mesh to the entire computational domain. In this case, within the region of the auxiliary mesh that covers the atomistic region, the magnetisation $\vectorn{M}$ in \eqref{eq:EquationsForPotential} is obtained by volume-averaging of the atomistic spin magnetic moments, while the atomistic demagnetising field $\vectorn{H}_{\mathrm{a},i}$ is obtained by interpolation of the continuum demagnetising field $\vectorn{H}_\mathrm{c}$.

It is also possible to have the force-based approach and FEM-discretised continuum. In this case, since the conservation of the total energy becomes irrelevant, the interpolation can be used, which means that mesh used for the solution of the equation for the demagnetising field can be arbitrary. The final combination of the energy-based approach and FDM-discretised continuum is somewhat strange and probably does not have a practical purpose, as the construction of error-free interface coupling is not straightforward, i.e. the specific interaction between each interface node and surrounding atoms the minimises the ghost-forces/torques must be derived.

\subsubsection{The scheme used in the numerical examples}

In the numerical examples of this paper, the continuum region is discretised using the FDM. The regions are coupled using a variant of the force-based coupling, modified to be used together with the implicit time-stepping.

To provide the boundary conditions for the atomistic region, padding atoms are constructed, see figure \ref{fig:scheme}. The solution at padding atoms is obtained by bilinear interpolation of the continuum solution with subsequent normalisation. The normalisation is introduced to preserve the length of spin magnetic moments.

To be able to evolve the continuum solution using an implicit time-stepping method, the continuum mesh is extended to the entire computational domain. The solution at the continuum nodes, which overlap with the atomistic region, is obtained by a normalised weighted average of the atomistic solution inside the box with side $\Delta x$ centred at the the node, where $\Delta x$ is the continuum mesh size. For all atoms inside the box, the weight is assigned as the area of the intersection of the box with side $a$ centred at the atom and the box with side $\Delta x$ centred at the node. The normalisation is introduced to preserve the nodal length of the vector field solution.

Furthermore, the auxiliary mesh for solving \eqref{eq:EquationsForPotential} is introduced. It coincides with the extended continuum mesh, which is discussed above. Since the nodes of the meshes coincide, the magnetisation $\vectorn{M}$ in \eqref{eq:EquationsForPotential} is taken to be equal to the continuum magnetisation.

To reduce the high-frequency wave-reflection from the atomistic-continuum interface, additional numerical damping is added to atoms close to the atomistic-continuum interface \cite{Poluektov2016, Poluektov2018}. This damping acts as a low-pass filter for the waves travelling from the atomistic region to the continuum, as the solution is ``attenuated'' to an average solution within a certain window. Due to a dispersive nature of the spin waves, the damping is non-linear and depends on time derivative of the solution. The analysis of the dynamics of the damping layer and the exact form of the modification can be found in \cite{Poluektov2016}.

Following the force-based coupling methodology, the time stepping is performed separately for the atomistic and the continuum regions. The implicit mid-point method \cite{dAquino2005} is used to solve the equations in time. Within a particular time step, the continuum region (extended to the entire computational domain) is solved first to obtain the current time-step values, which includes the solution of the equation for the demagnetisation field. This gives the solution at the padding atoms at the current time step. The atomistic region is subsequently solved using the padding atoms as boundary conditions and the demagnetisation field at the current time step. Finally, the solution at the continuum nodes, which overlap with the atomistic region, is overwritten by volume-averaging of the atomistic solution.

\subsection{Heterogeneous multiscale methods}
\label{subsec:HMM}

Recently, an HMM approach has been formulated in application to multiscale problems arising in micromagnetism. First, in \cite{Arjmand_etal_2016}, a multiscale method has been proposed to simulate the coarse-scale dynamics of a chain of atomistic spins. The atomistic spins were subjected to a high-frequency external field and a mathematical investigation of the convergence rates in relation to the coupling/upscaling errors, originating from a micro-macro coupling, was given for a simplified setting of a single spin.

At finite temperatures, the atomistic LLG equation \eqref{eq:ASD_LLG} also includes a white noise term. The noise term results in fluctuations of the magnetic moment vectors $\vectorn{m}_i$. The macroscopic quantities of interests, in this case, are the expected values of the local averages (in space and time) of the magnetic moments. By taking the inner product of the equation \eqref{eq:ASD_LLG} with $\vectorn{m}_i$, it is easy to see that the length of each individual moment $\vectorn{m}_i(t)$ is equal to $\left| \vectorn{m}_i(t) \right| = 1$, $\forall t$. However, due to the thermal fluctuations, the statistical averages acquire reduced lengths, i.e. $\left| E[\vectorn{m}_i](t) \right| < 1$. This has been the major reason for the development of the finite-temperature HMM-based model \cite{APK_Temp_2017}. In both HMM-based algorithms, for zero and for non-zero temperature, the modelling of the long-range interactions has not been considered, as the main ambition has been to model the local terms and the temperature effects accurately, in \cite{Arjmand_etal_2016} and \cite{APK_Temp_2017}, respectively. In subsections \ref{Sec:HMM_Zero_Temprature} and \ref{Sec:HMM_NonZero_Temprature} of this paper, an extension of the algorithms from \cite{Arjmand_etal_2016,APK_Temp_2017} is presented, after introducing the mathematical tools and notations in subsection \ref{Sec:AveragingKernels}. In particular, it is demonstrated that the micro problems associated with both multiscale methods must be modified in a suitable way to capture the correct macroscopic dynamics in the presence of the long-range interactions. 

\subsubsection{Averaging kernels} 
\label{Sec:AveragingKernels}

In this subsection, the basic mathematical tools and notations for the HMM algorithms in subsections \ref{Sec:HMM_Zero_Temprature} and \ref{Sec:HMM_NonZero_Temprature} are introduced. The HMM algorithms developed in \cite{Arjmand_etal_2016} and \cite{APK_Temp_2017} are based on the notion of upscaling, where a local average of small scale features in the atomistic solution \eqref{eq:ASD_LLG} is computed and used in a macroscale model. The local averaging takes place in small domains in space and time. In practice, the spatial size of the averaging is comparable to the size of a few interatomic distances, i.e. $\eta = m a$, $m  \in \mathbb{Z}^{+}$. The temporal averaging, however, takes place on a domain of size $\tau  = O(\e)$, where $\e$  is a time scale, at which the microscopic dynamics undergoes some variations. For averaging, the space $\mathbb{K}^{p,q}$ of averaging kernels (weight functions) is introduced. The space $\mathbb{K}^{p,q}$ consists of functions $K$, which have compact support in $[-1/2,1/2]$, and
\begin{itemize}
\item{$K$ is symmetric, i.e., $K(t) = K(-t)$}
\item{$K^{(q+1)}(t) \in BV(\mathbb{R})$, where $BV$ is the space of functions with bounded variations in $\mathbb{R}$}
\item{$K$ has $p$ vanishing moments, i.e.
\begin{equation*}
  \int_{\mathbb{R}} K(t) t^r \; \mathrm{d}t =
    \begin{cases} 1, & r=0 , \\
    0, & 0<r\leq p .
  \end{cases}
\end{equation*}
}
\end{itemize}
Applying a kernel $K \in \mathbb{K}^{p,q}$ to an $\e$-periodic function $f^{\e}(t) = f(t/\e)$, where $f$ is $1$-periodic, with an average defined as $\bar{f} := \int_{0}^{1} f(s) \; \mathrm{d}s$ results in the following arbitrarily high convergence rates, see e.g. \cite{Arjmand_Runborg2016a,Arjmand_Runborg2014},
\begin{equation}
  \left| \left( K_{\tau} \ast f^{\e} \right)(0) - \bar{f} \right| \leq C \left(  \dfrac{\e}{\tau} \right)^{q+2}, \quad \tau > \e,
  \label{eq:Kernel_Conv_Rate}
\end{equation}
where $K_{\tau}(\cdot):=\frac{1}{\tau} K(\cdot/\tau)$ is a scaled kernel, $q$ is the smoothness parameter associated with $K$, and 
\begin{equation*}
  \left( K_{\tau} \ast f^{\e} \right)(t) := \int_{t-\tau/2}^{t+\tau/2} K_{\tau}(s-t) f^{\e}(s) \; \mathrm{d}s .
\end{equation*}
Note that a constant kernel belongs to the space $\mathbb{K}^{1,-1}$, i.e. $q=-1$, and, therefore, the corresponding error becomes $O(\e/\tau)$, in view of the estimate \eqref{eq:Kernel_Conv_Rate}. In general, smoother the kernel $K$ (higher $q$), higher the convergence rates become. Moreover, if $\tau$ is an exact integer multiple of $\e$, then the constant $C$ in the above estimate is zero and the averaging is exact. For a numerical verification of the convergence rate in the above estimate and more general results for non-periodic integrands, see \cite{Arjmand_Runborg2016a,Arjmand_Runborg2014,Engquist_Tsai_2005}.

These local averaging kernels will be used in the description of the HMM algorithms below. 

\subsubsection{HMM at zero temperature}
\label{Sec:HMM_Zero_Temprature}

To present the numerical method, a 1D case is considered. The extension of the algorithm to higher dimensions is self-evident, and skipped in the exposition. First, it is assumed that the full atomistic system consists of $N = (r+\ell)L +1$ number of magnetic moments that are located on a set of discrete points in 1D, i.e. $\{ x_i  = i a \}_{i=0}^{(r+\ell)L}$, where $a$ represents the interatomic distance, $r \in \mathbb{Z}^{+}$ and $\ell \in \mathbb{N}$ are two non-negative integers. Moreover, the magnetic moments are supplied with periodic boundary conditions (BCs) and are the solutions of the atomistic LLG model \eqref{eq:ASD_LLG} with $ \vectorn{H}_i =  \left(   \sum_{j}  J_{ij} \vectorn{m}_{j} + \vectorn{H}^{\e}_{\mathrm{e},i}(t) + \vectorn{H}_{\mathrm{a},i} \right)$, where $\vectorn{H}^{\e}_{\mathrm{e},i}(t)$ is a high-frequency external field oscillating with the wavelength $\e$ in time. The index $i$ in $\vectorn{H}_{\mathrm{e},i}^{\e}$ is to allow for spatially non-uniform external fields. For the exchange coefficient a nearest-neighbour interaction is assumed, i.e. 
\begin{equation*}
J_{ij}  = \begin{cases} J, & i=j , \\
0, & \text{otherwise.}
\end{cases}
\end{equation*}

The macroscopic variable is defined as the local average of $(2 r + 1)$ microscopic magnetic moments. To define the macro variable, assume that the coarse grid is given by $\{ X_{I}  = I (r+ \ell)  a\}_{I=0}^{L}$ with $L \ll N$, implying much fewer degrees of freedom in comparison to a full atomistic model. The macroscopic magnetisation $\vectorn{M}_I$ at a point $X_I$ is defined as 
\begin{align} \label{DefnMacroM_eqn}
\vectorn{M}_I(t) &= a \sum_{j=-r}^{r} K_{\eta}(x_{I(r+ \ell)+j}   -   x_{I(r+\ell)})  \left(  K_{\tau} \ast \vectorn{m}_{I(r+\ell)+j} \right)(t) \\ &=:  \left( \mathcal{K}_{\tau,\eta} \ast \vectorn{m} \right)(X_I,t), \nonumber
\end{align}
where $\eta = (2 r + 1) a$ is the size of the local spatial averaging domain and $\tau > \e$ is that of a temporal averaging. From formula \eqref{DefnMacroM_eqn}, it is evident that between two consecutive macroscopic points, a total number of $\ell$ magnetic moments is skipped while averaging.

{\bf Macro model.} The macro model takes the form
\begin{align} \label{MacroProblemChain_eqn}
  \dfrac{\dif }{\dif t} \vectorn{M}_I(t) &= -\vectorn{F}_I(t,\vectorn{M}_{\tilde{I}}) - 
    \dfrac{\alpha_{\mathrm{L}}}{\beta_{\mathrm{L}}} \vectorn{M}_I \times \vectorn{F}_I(t,\vectorn{M}_{\tilde{I}}), \quad t \in (0,T], \\
  \vectorn{M}_I(0) &= \left( \mathcal{K}_{\tau,\eta} \ast \vectorn{m}\right)(X_I, 0), \nonumber
\end{align}
where $\tilde{I} = \{I-1,I,I+1 \}$. 

{\bf Micro model.} To close the macro problem, $\vectorn{F}_I(t^{*},\vectorn{M}_{\tilde{I}})$ must be computed. To do this, a set of coupled non-linear ODEs for  $\vectorn{m}_{I r^{\prime} + j}(t)$, where $r^{\prime} = r + \ell$, $t \in \mathcal{I}^{\pm}_{\tau}$, $\mathcal{I}^{+}_{\tau} = t^{*} + [0,\tau/2]$, $\mathcal{I}^{-}_{\tau} = t^{*} + [-\tau/2,0]$, with $\tau/2 > \e$, and $j =-r+1,\ldots,r-1 $ is solved, i.e. 
\begin{align} \label{MicroProblemChain_eqn}
  \dfrac{\dif}{\dif t} \vectorn{m}_{Ir^{\prime} + j} (t) &= \beta_{\mathrm{L}} \vectorn{m}_{I r^{\prime} + j} \times
    \left( \sum_k{J_{{Ir^{\prime}  + j}, k} \vectorn{m}_k} + \vectorn{H}^{\e}_{\mathrm{e},Ir^{\prime} + j}(t) +
    \vectorn{H}_{\mathrm{c}}(\vectorn{M}, x_{Ir^{\prime}  + j})  \right) , \\
  \vectorn{m}_{Ir^{\prime} + j}(t^{*})  &= \hat{\vectorn{M}}(x_{Ir^{\prime} + j})  \nonumber \\
  \vectorn{m}_{Ir^{\prime} - r}(t) &= \hat{\vectorn{M}}(x_{Ir^{\prime} - r}), \quad
  \vectorn{m}_{Ir^{\prime} + r}(t) = \hat{\vectorn{M}}(x_{Ir^{\prime} + r}), \nonumber
\end{align}
where $\hat{\vectorn{M}} = \pi_2 \vectorn{M}/|\pi_2 \vectorn{M} |$ is obtained by a normalised second order polynomial interpolation of the macroscopic solutions $\vectorn{M}_{\tilde{I}}$ for $\tilde{I} = I-1,I,I+1$. 

{\bf Upscaling.} The last step is to upscale the quantity $\vectorn{F}_I(t^{*},\vectorn{M}_{\tilde{I}})$ in \eqref{MacroProblemChain_eqn} by 
\begin{equation}\label{UpscalingChain_eqn}
  \vectorn{F}_{I}(t^{*},\vectorn{M}_{\tilde{I}}) = \left( \mathcal{K}_{\tau,\eta} \ast \dfrac{\dif}{\dif t} \vectorn{m} \right)\left( X_I,t^{*} \right). 
\end{equation}

Note that instead of a full atomistic simulation over the entire computational domain, a fewer number of atoms ($2r+1$ atoms) are coupled together in the micro problem \eqref{MicroProblemChain_eqn}. Moreover, the boundary atoms and the initial data of the micro problem are forced to be equal to the coarse-scale variables, to synchronise the microscopic model with macro variables. In \eqref{MicroProblemChain_eqn}, the quantity $\vectorn{H}_{\mathrm{c}}$ is computed by solving the equation \eqref{eq:EquationsForPotential} on the macroscopic grid. Moreover, in the computation of $\vectorn{H}_{\mathrm{c}}$, the macro solutions $\vectorn{M}$ are used in the right hand side of \eqref{eq:EquationsForPotential}. It is worth mentioning that the damping term is not included in the micro model, since it is modelled at the macroscopic level, equation \eqref{MacroProblemChain_eqn}. This is similar to the HMM algorithm from \citep{Arjmand_etal_2016}, where the convergence of the macroscopic solutions to the exact coarse-scale solutions has been proved in the absence of the long-range field. The main novelty of the current algorithm is the replacement of the microscopic long-range interaction field with the continuum long-range field $\vectorn{H}_{\mathrm{c}}$, which can be efficiently approximated using a standard finite difference/element method on the macroscopic domain. This approach leads to a tremendous gain in computational cost due to the fact that the atomistic computation of the long-range field is avoided, which would otherwise require the atoms in a given microscopic domain to communicate with the atoms located in neighbouring microscopic domains over a large macroscopic geometry.    

\begin{Remark}
In principle, the macro, \eqref{MacroProblemChain_eqn}, and the micro, \eqref{MicroProblemChain_eqn}, problems can be discretised by any convergent time-stepping method. But if certain discrete conservation properties are required, a special care must be given to the choice of the method. The particular choice of the numerical methods, used for the simulations in this paper, can be found in the numerical results section; see also  \cite{Garcia2007} and the references therein for a review about time stepping methods in micromagnetism. 
\end{Remark}

\begin{Remark}
In general, the macroscopic quantities are much smoother in time and space, as they do not `see' the variations at atomic scales. Hence, in computations, the macroscopic model \eqref{MacroProblemChain_eqn} is discretised using a time step $\Delta t$, which is much larger than a time step $\delta t$ used for a discretisation of the micro model \eqref{MicroProblemChain_eqn}.
\end{Remark}

\subsubsection{HMM at non-zero temperature}
\label{Sec:HMM_NonZero_Temprature}

An extension of the zero temperature algorithm from \cite{Arjmand_etal_2016} to non-zero temperature, was introduced in \cite{APK_Temp_2017}. Modelling the long-range interactions requires yet another set of modifications to the algorithm from \cite{APK_Temp_2017}. To describe these modifications, let 
\begin{align}
  \label{eq:DeterministicOscillatoryH}
  &\vectorn{H}_{\mathrm{det},i} = \left(  \sum_{j} J_{ij} \vectorn{m}_{j} \right) + 
    \tensor{K}_{\mathrm{a}} \cdot \vectorn{m}_i + \vectorn{H}_{\mathrm{e},i}^{\e}(t) , \\
  \label{eq:StochasticH}
  &\vectorn{H}_{\mathrm{sto},i}(t,t^{\ast}) = \left(  \sum_{j} J_{ij} \vectorn{m}_{j} \right) + 
    \tensor{K}_{\mathrm{a}} \cdot \vectorn{m}_i + \left( K_{\tau} \ast\vectorn{H}_{\mathrm{e},i}^{\e} \right)(t^{\ast}) + \vectorn{h}_i(t)
\end{align}
be adopted. Note the differences between  $\vectorn{H}_{\mathrm{det},i}^{\e}$ and $\vectorn{H}_{\mathrm{sto},i}$. The term $\vectorn{H}_{\mathrm{det},i}^{\e}$ is deterministic but oscillatory, while $\vectorn{H}_{\mathrm{sto},i}$ is stochastic  and includes the filtered external field $\left( K_{\tau} \ast\vectorn{H}_{\mathrm{e},i}^{\e} \right)(t^{\ast}) $. The superscript $\e$ in the term $\vectorn{H}^{\e}_{\mathrm{e}}$  is to denote that the external field has high frequency variations. 

The model at non-zero temperature requires a modification of the zero temperature model. In particular, an additional step is needed to capture the reduction in the length, which arises from taking statistical averages of atomic moments. The precise algorithm (in the presence of the long-range interactions) is given below.

{\bf Macro model.} With a slight deviation to the algorithm at zero temperature, the macro model at nonzero temperatures is given by 
\begin{align}
  \label{eq:Macromodel}
  s_I(t) \dfrac{\mathrm{d}}{\mathrm{d}t} \vectorn{M}_{I}(t) &= -\vectorn{M}_I(t) \dfrac{\mathrm{d}}{\mathrm{d}t} s_{I}(t) -
    \vectorn{F}_{I}(t,s_{\tilde{I}}\vectorn{M}_{\tilde{I}}) - \dfrac{\alpha_\mathrm{L}}{\beta_\mathrm{L}} s_I(t) \vectorn{M}_{I} (t) \times
    \vectorn{F}_{I}(t, s_{\tilde{I}}\vectorn{M}_{\tilde{I}}), \nonumber \\
  \vectorn{M}_I(0) &= \left( \mathcal{K}_{\tau,\eta} \ast \vectorn{m}\right)(X_I, 0),
\end{align}
where $\tilde{I}:=\left\{ I-1,I,I+1\right\}$, $\vectorn{F}_{I}$ is the missing data in the model and $s_I(t) \vectorn{M}_I(t)$ is the ultimate macro solution, which models the coarse-scale dynamics. In particular, $\vectorn{M}_I(t)$ has unit length, up to an upscaling error, and represents the direction and $s_I(t) \leq  1$, which is computed below, accounts for the reduction in the magnetisation length.

{\bf Micro model.} To compute $\vectorn{F}_I(t^{*},s_{\tilde{I}}\vectorn{M}_{\tilde{I}})$, first the micro problem
\begin{align}
  \label{eq_Micromodel}
  \dfrac{\mathrm{d}}{\mathrm{d}t} \vectorn{m}_{I r^{\prime}+j}(t)  &= \beta_\mathrm{L} \vectorn{m}_{Ir^{\prime}+j}(t) \times
    \left( \vectorn{H}_{\mathrm{det},Ir^{\prime}+j}(t) +
    \vectorn{H}_{\mathrm{c}}(s_{\tilde{I}} \vectorn{M}_{\tilde{I}}, x_{Ir^{\prime}+j})\right), \\
  \vectorn{m}_{Ir^{\prime}+j}(t^{*}) &= \hat{\vectorn{M}}(x_{Ir^{\prime}+j}), \nonumber \\
  \vectorn{m}_{Ir^{\prime} - r}(t) &=  \hat{\vectorn{M}}(x_{Ir^{\prime} - r}), \quad
  \vectorn{m}_{Ir^{\prime}+r}(t) = \hat{\vectorn{M}}(x_{Ir^{\prime} + r}), \nonumber
\end{align}
needs to be solved. Here $t \in \mathcal{I}^{\pm}_{\tau}$, $\mathcal{I}^{+}_{\tau} := t^{*} + [0,\tau/2]$, $\mathcal{I}^{-}_{\tau}:= t^{*} + [-\tau/2,0]$, and $r^{\prime} = r + \ell$. The index $j$ is in the range $j =-r+1,\ldots,r-1$. Moreover, the final microscopic time $\tau/2$ satisfies $\tau/2 > \e$ and $\hat{\vectorn{M}}(x) = \pi_2 (s_{\tilde{I}}\vectorn{M}_{\tilde{I}})/|\pi_2 (s_{\tilde{I}}\vectorn{M}_{\tilde{I}})|(x)$ denotes the normalised second order polynomial interpolation of the macroscopic solutions. Finally, the term $\vectorn{H}_{\mathrm{c}}$ is computed similarly as in the zero temperature HMM algorithm in the previous subsection.

{\bf Upscaling.} The quantity $\vectorn{F}_I(t^{*},s_{\tilde{I}}\vectorn{M}_{\tilde{I}})$ is computed by
\begin{align}
  \label{eq:Upscaling}
  \vectorn{F}_{I}(t^{*},s_{\tilde{I}}\vectorn{M}_{\tilde{I}}) = \left( \mathcal{K}_{\tau,\eta} \ast \dfrac{\dif}{\dif t} \vectorn{m} \right)\left(X_I,t^{*}\right)
\end{align}
where $\eta = (2 r+1)a$.

{\bf Computation of $s_I(t)$.} In the final step, the quantity $s_{I}(t^{*})$ is computed by solving the following stochastic LLG equation for $j = -r+1,\ldots,r-1$ and $t \in [t^{*},t^{*}+\tau_\mathrm{f}]$ 
\begin{align}
  \label{eq:LengthScaling}
  \dfrac{\mathrm{d}}{\mathrm{d}t} \vectorn{m}_{Ir^{\prime}+j}(t) &= -\beta_\mathrm{L} \vectorn{m}_{Ir^{\prime}+j}(t) \times
    \left( \vectorn{H}_{\mathrm{sto},Ir^{\prime}+j}(t;t^{*}) + \vectorn{H}_{\mathrm{c}}(s_{\tilde{I}} \vectorn{M}_{\tilde{I}},
    x_{Ir^{\prime}+j}) \right) \nonumber \\
  &\hphantom{=} - \alpha_\mathrm{L} \vectorn{m}_{Ir^{\prime}+j} \times (\vectorn{m}_{Ir^{\prime}+j} \times
    \left( \vectorn{H}_{\mathrm{sto},Ir^{\prime}+j}(t;t^{*})) + \vectorn{H}_{\mathrm{c}}(s_{\tilde{I}} \vectorn{M}_{\tilde{I}},
    x_{Ir^{\prime}+j}) \right) , \\ 
  \vectorn{m}_{Ir^{\prime}+j}(t^{*}) &= \hat{\vectorn{M}}(x_{Ir^{\prime}+j}), \nonumber \\
  \vectorn{m}_{Ir^{\prime} - r}(t) &= \hat{\vectorn{M}}(x_{Ir^{\prime} - r}), \quad
    \vectorn{m}_{Ir^{\prime} + r}(t) = \hat{\vectorn{M}}(x_{Ir^{\prime} + r}), \nonumber
\end{align}
where $\tau_\mathrm{f} > \tau_\mathrm{r}$, and $\tau_\mathrm{r}$ is the time it takes to reach the thermal equilibrium, and $\hat{\vectorn{M}}$ is defined similarly as in the micro problem.
Then, with $\eta = (2 r + 1)a$, the following is computed
\begin{align}
  \label{eq:LengthComput}
  s_{I}(t^{*})  = \left| a \sum_{j=-r}^{r} K_{\eta}(x_{Ir^{\prime}+j}-x_{Ir^{\prime}}) \left( \dfrac{1}{\tau_\mathrm{f}-\tau_\mathrm{r}} \int_{\tau_\mathrm{r}}^{\tau_\mathrm{f}} \vectorn{m}_{Ir^{\prime}+j} (t) \;\mathrm{d}t \right) \right|. 
\end{align}

Note that in the zero and the non-zero temperature algorithms, the long-range continuum field appears in the micro problem. Moreover, for the non-zero temperature HMM, it is necessary to include the long-range continuum field in the length scaling procedure, equation \eqref{eq:LengthScaling}, as well. This is due to the fact that the magnetisation length is also influenced by the long-range field, see e.g. \cite{Aharoni1996} for a mathematical motivation.

\section{Computational examples}
\label{Sec_NumericalResults}

\subsection{Partitioned-domain example: Domain wall kinetics in a 2D structure}

The advantages of the partitioned-domain mutiscale technique are revealed in cases when the atomistic resolution is required locally, while the rest of the computational domain is homogeneous and can be approximated with sufficient accuracy by the continuum model. One such example is the domain wall kinetics in a material with local defects. In \cite{Poluektov2018}, such a problem was considered and the performance and the advantages of the partitioned-domain multiscale technique were demonstrated. However, in \cite{Poluektov2018}, the domain wall was created using only the exchange and the anisotropy terms in the LLG equation. In this paper, a similar example is considered, however, in which the domain wall is created by the exchange and the demagnetisation terms. The field-induced movement of the domain wall in the presence of a void in the magnetic structure is investigated. From the physical point of view, the void in the material can correspond to a micro-crack of the sample or to impurity atoms.

The domain wall in a material with the 2D $\left(111\right)$ fcc stacking of atoms is considered. The material contains a hexagonal void with the side of $n a$, where $a$ is the lattice spacing and $n$ is ranging from $3$ to $6$. The major effect that is observed in the simulations, is the blocking of the domain wall by the void of size $6a$, while for the size of the void up to and including $5a$, the domain wall is only slowed down by the void.

\subsubsection{Computational setup and model parameters}

Since all quantities are considered to be dimensionless, $\beta_\mathrm{L} = 1$ was used. The damping was selected to be $\alpha_\mathrm{L} = 0.1$. Atoms were selected to be arranged according to 2D $\left(111\right)$ fcc stacking. Lattice spacing was taken to be $a = 1/64$. The exchange coefficients were taken to be $J_{ij}/\mu = (2/3)a^{-2}$, which gives the continuum exchange tensor $\tensor{A}_\mathrm{e} = A_\mathrm{e} \tensor{I}$ with $A_\mathrm{e}/\mu = 1$, where $\tensor{I}$ is the identity tensor. The anisotropy term was not taken into account either, $\tensor{K}_\mathrm{a} = 0$. The coefficient that defines the magnitude of the demagnetisation field was taken to be $c_\mathrm{L} = 2 \pi^2$, which gives the approximate width of the domain wall $w_\mathrm{D} = 1$. The external field was applied in the $x$-direction, $\vectorn{H}_\mathrm{e} = H_x \vectorn{e}_x$, where $H_x = 5$ was assumed.

The width and the height of the computational region were taken to be $x_\mathrm{L} = 4$ and $y_\mathrm{L} = 1$, respectively. Neumann boundary conditions were used at $x = 0$, $x = x_\mathrm{L}$, $y = 0$, and $y = y_\mathrm{L}$ for both LLG and magnetostatic equations. The computational region was partitioned into the atomistic and the continuum subregions. The continuum discretisation step was taken to be $\Delta x = 4a$. The atomistic subregion was located in the centre of the computational region. The width and the hight of the atomistic subregion were taken to be $x_\mathrm{A} = y_\mathrm{A} = 48a = 0.75$. The atomistic subregion contained a hexagonal void with a side of $n a$, where $n \in \left\lbrace 3,4,5,6 \right\rbrace$, the centre of which was located at $x = 2$ and $y = 0.5$. Time step $\Delta t = 10^{-2}$ was used.

Within the atomistic region, the behaviour of atoms close to the interface was modified with the additional numerical damping. Parameters of the damping region, the optimal values of which depend on the width of the damping region and the difference between discretisations, were taken from \cite{Poluektov2018}, where the same difference between the discretisations of the regions was used. The damping strength and the width of the averaging window was taken to be $g_\mathrm{D} = 625$, $s_\mathrm{A} = 3 \Delta x$. The width of the damping band was selected to be $16a$, as the large width of the damping band ensures that the atomistic solution is not contaminated by wave reflections.

The following initial conditions for the domain wall were used:
\begin{align*}
  &\vectorn{M} = \vectorn{e}_y \sin \theta + \vectorn{e}_z \cos \theta , \\
  &\theta = \operatorname{arcsin} \left( \tanh \left( \pi \sqrt{2}
    \left( x - x_0 \right) \right) \right) + \frac{\pi}{2} ,
\end{align*}
where $x_0$ is the position of the centre of the domain wall. The initial position of the centre of the domain wall was taken to be $x_0 = 1$.

Since the domain wall moves during the simulation, to analyse the results, it is important to obtain from the simulations the exact position of the centre of the domain wall as a function of time. The domain wall centre is defined as the curve, along which $M_z=0$. Since the computational solution, $\vectorn{M}$, is defined at the grid points, an auxiliary quantity $\zeta = \operatorname{arccos} \left(M_z\right) - \pi/2$ is calculated at each grid point and linearly interpolated between the grid points. Thus, for each $y = y_0$, the domain wall centre along $x$-axis, $x_0$, is found by solving $\zeta = 0$.

\subsubsection{Results}

\begin{figure}
  \begin{center}
    \includegraphics[scale=0.7]{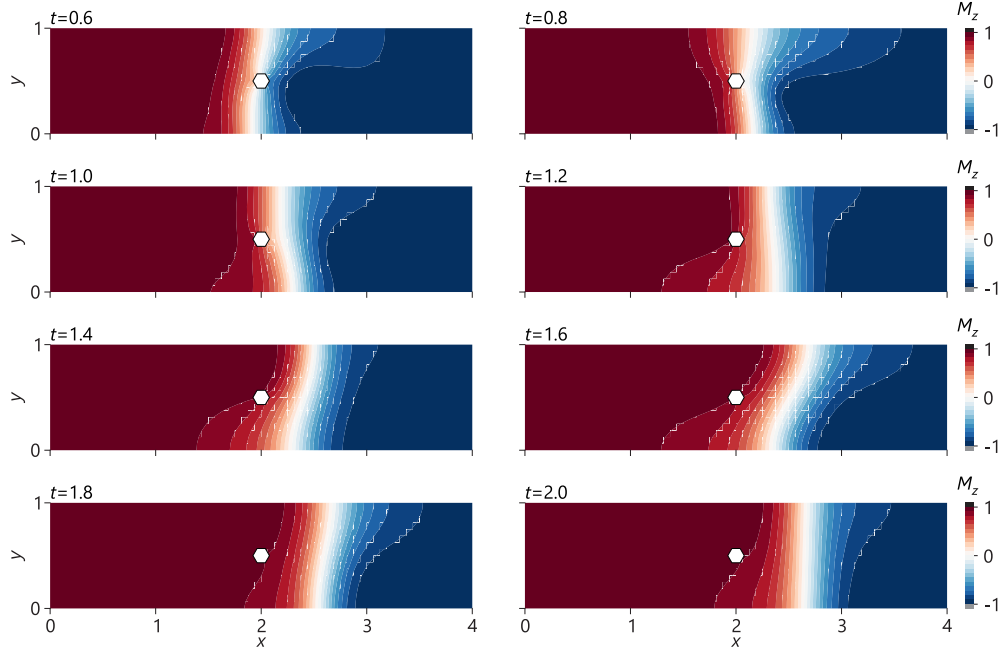}
  \end{center}
  \caption{The distribution of $M_z$ component of magnetisation in the computational domain at different simulation times.}
  \label{fig:res_field}
\end{figure}

In figure \ref{fig:res_field}, the field plot of the $z$-component of magnetisation for the case of $n=5$ is illustrated. It can be seen that when the domain wall approaches the void, the thickness of the domain wall decreases locally. The region of the domain wall that is located in the upper half of the 2D plate slows down, while the the lower part of the domain wall moves past the void ($t=0.8$ and $t=1.0$). Afterwards, the upper part of the domain wall accelerates and overtakes the lower part ($t=1.4$ and $t=1.6$). Finally, an equilibration process is observed ($t=1.8$ and $t=2.0$). Thus, the void causes oscillations in the structure of the domain wall. Moreover, in the field plots, it can be seen that the void acts as a ``gradient concentrator'', i.e. the region around the void creates higher gradient in comparison to the regions further from the void.

From the physical point of view, the energetically preferred states of spin magnetic moments are approximate alignments of spins either along $\vectorn{e}_z$ or opposite to $\vectorn{e}_z$. The intermediate states of spins, which correspond to the domain wall, have higher energy. When the domain wall passes through the void, a region of the wall is absent (due to the void) and, thus, a number of ``high-energy'' spins is absent. Therefore, to minimise the total energy, the preferred states of the spin system are such that the void covers spins with the highest energy locally. This explains why the domain wall and the gradient lines in figure \ref{fig:res_field} tend to stick to the void.

\begin{figure}
  \begin{center}
    \includegraphics[scale=0.8]{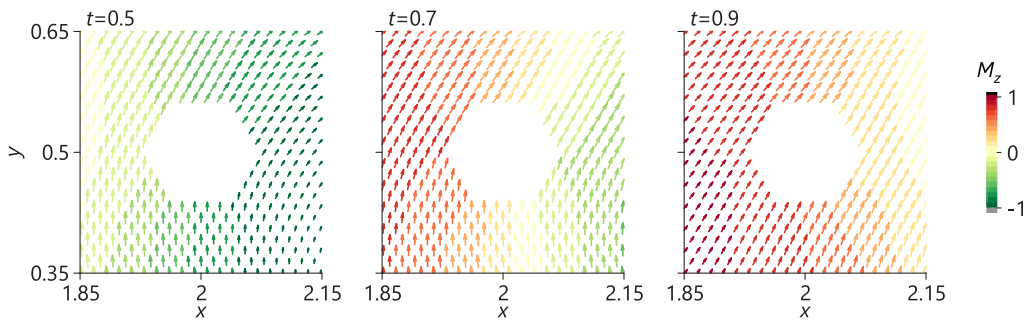}
  \end{center}
  \caption{The magnetic moments of the individual atoms in the region around the void at different simulation times. The magnetic moments are shown using the vector field and are projected onto $xy$-plane, i.e. spins that are parallel to $\vectorn{e}_z$ are dots, while spins that are lying in $xy$-plane have the length of $a/2$.}
  \label{fig:res_vect}
\end{figure}

The states of spin magnetic moments around the void when the domain wall passes the void are shown in figure \ref{fig:res_vect}. Although the initial structure of the domain wall is of the Bloch-type, i.e. the spins have zero $M_x$ component, it is clearly seen that as the domain wall interacts with the void, spins acquire a non-zero $M_x$ component. This is the mechanism by which the domain wall is slowed down by the void in this example. Since the external field is aligned with $\vectorn{e}_x$, the torque that acts on the spins with significant $M_x$ component is small, which decreases the angular velocity of the spins in the centre of the domain wall and the thereby leads to the decrease of the speed of the domain wall propagation. Moreover, at $t=0.7$, it is seen that the upper part of the domain wall is inclined, as opposed to the lower part of the domain wall, which is vertical. This corresponds to the moment when the upper part of the domain wall moves faster than the lower part.

\begin{figure}
  \begin{center}
    \includegraphics[scale=0.8]{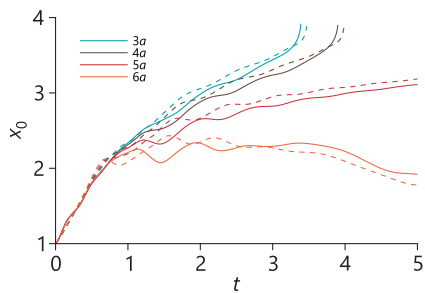}
  \end{center}
  \caption{The dependencies of the positions of the domain wall centre at $y = 0.25$ (solid lines) and at $y = 0.75$ (dash lines) on time for different sizes of the void, which are indicated in the legend.}
  \label{fig:res_pos}
\end{figure}

To understand the dependence of the kinetics of the domain wall on the size of the void, the position of the domain wall centre was calculated for each time step. The dependencies of the positions of the domain wall centre at $y = 0.25$ and at $y = 0.75$ on time are shown in figure \ref{fig:res_pos}. The void is located at $x=2$ and it is clearly seen that when the domain wall passes the void, the velocity drops. The decrease of the velocity non-linearly depends on the size of the void. For the void of size $3a$ the decrease of the velocity is relatively small, while for $5a$ it is already significant. For the void of the size of $6a$, the position of the domain wall stays below $2.5$, i.e. the domain wall does not move past the void and stays at $x \approx 2$.

In figure \ref{fig:res_pos}, it is also seen that there is a difference between the positions at $y = 0.25$ and at $y = 0.75$. Increases and subsequent decreases of the positions, where the lower part of the domain wall overtakes and then falls behind the upper part of the domain wall, are the oscillations created by the void. The magnitude of these oscillations also depends on the size of the void and increases with the increase of the size (in the considered example, the highest oscillations in the domain wall position are observed for the void of size $6a$).

\subsection{Partitioned-domain example: Multiscale modelling error, 1D example}

The proposed multiscale technique obviously has a modelling error, which is the error due to the representation of the demagnetisation field using the discretised continuum model. This error is proportional to $c_\mathrm{L}$, which can be understood as the parameter defining the magnitude of the demagnetisation field. This is demonstrated in this section using an example of a 1D domain wall moving form the continuum region into the atomistic region.

As was shown above, in the 1D case, $\vectorn{H}_\mathrm{c} = -c_\mathrm{L} M_x \vectorn{e}_x$. This means that the demagnetisation field acts similar to the anisotropy, but with the negative sign. Moreover, due to the structure of the LLG equation \eqref{eq:C_LLG}, $C\vectorn{M}$ can be added to effective field $\vectorn{H}$, where $C$ is an arbitrary constant, without influencing the solution of the LLG equation. This is used in the example below.

\subsubsection{Computational setup and model parameters}

The same $\beta_\mathrm{L}$ and $\alpha_\mathrm{L}$ as in the example above were used. Atoms were selected to be arranged in a 1D chain. Lattice spacing was taken to be $a = 1/64$. The exchange coefficients were taken to be $J_{ij}/\mu = a^{-2}$, which gives the continuum exchange parameter $A_\mathrm{e}/\mu = 1$. Biaxial anisotropy was used,
\begin{equation*}
  \tensor{K}_\mathrm{a} = \left( 2\pi^2 - c_\mathrm{L} \right) \mu \vectorn{e}_y \vectorn{e}_y -
  \left( 1 + c_\mathrm{L} \right) \mu \vectorn{e}_z \vectorn{e}_z .
\end{equation*}
The introduction of parameter $c_\mathrm{L}$ into the anisotropy is explained below. The coefficient that defines the magnitude of the demagnetisation field, $c_\mathrm{L}$ was varied from $0.4$ to $12.8$. These parameters give the approximate width of the domain wall $w_\mathrm{D} = 1$. The external field was applied in $z$-direction, $\vectorn{H}_\mathrm{e} = H_z \vectorn{e}_z$, where $H_z = 1$ was taken.

It can be seen that in the above presented setup, the exact continuum solution of the moving domain wall should not be affected by parameter $c_\mathrm{L}$, because
\begin{multline*}
  \frac{1}{\mu}\tensor{K}_\mathrm{a} \cdot \vectorn{M} + \vectorn{H}_\mathrm{c} = 
  2\pi^2 M_y \vectorn{e}_y - c_\mathrm{L} M_y \vectorn{e}_y - M_z \vectorn{e}_z - c_\mathrm{L} M_z \vectorn{e}_z
  -c_\mathrm{L} M_x \vectorn{e}_x = \\
  = 2\pi^2 M_y \vectorn{e}_y - M_z \vectorn{e}_z - c_\mathrm{L} \vectorn{M} 
\end{multline*}
and addition of $c_\mathrm{L}\vectorn{M}$ to $\vectorn{H}$ does not affect the solution of the LLG equation. However, in the computational solution, there will be a numerical error due to different treatments of the demagnetisation field and the other interactions. Also, it should be noted that negative anisotropy in $\vectorn{e}_z$-direction was chosen. Although this is not a realistic scenario, selection of a negative coefficient is mathematically allowed. It allows creating a setup, for which the exact continuum solution is independent of $c_\mathrm{L}$.

The length of the computational region was taken to be $x_\mathrm{L} = 6$. Neumann boundary conditions were used at $x = 0$, $x = x_\mathrm{L}$ for the LLG equation. The 1D analytical solution of the magnetostatic equation, $\vectorn{H}_\mathrm{c} = -c_\mathrm{L} M_x \vectorn{e}_x$, was used at the continuum scale to isolate the modelling error, i.e. not to introduce an error due to numerical solution of \eqref{eq:C_HLR} and \eqref{eq:EquationsForPotential}. The computational region was partitioned into the atomistic and the continuum subregions. The continuum discretisation step was taken to be $\Delta x = 4a$. The atomistic subregion was located in the centre of the computational region. The length of the atomistic subregion was taken to be $x_\mathrm{A} = 128a$. Time step $\Delta t = 10^{-2}$ was used. Same $g_\mathrm{D}$, $s_\mathrm{A}$ and the width of the damping band as in example above are used.

The following initial conditions for the domain wall were used:
\begin{align*}
  &\vectorn{M} = \vectorn{e}_x \sin \theta \cos \varphi +
    \vectorn{e}_y \cos \theta \cos \varphi + \vectorn{e}_z \sin \varphi , \\
  &\theta = \operatorname{arcsin} \left( \tanh \left( \pi \sqrt{2}
    \left( x - x_0 \right) \right) \right) + \frac{\pi}{2} , \\
  &\varphi = \operatorname{arcsin} \frac{1}{2\pi^2+1} ,
\end{align*}
where $x_0$ is the position of the centre of the domain wall. The initial position of the centre of the domain wall was taken to be $x_0 = 1.5$.

The simulations were run until $t_\mathrm{end} = 4$ and compared at that point. This roughly corresponded to the domain wall being in the centre of the atomistic region. The reference solution with $c_\mathrm{L} = 0$ was used. The solution in the atomistic region was used for the error calculation. The error was defined as the $L^1$-norm of the difference between solutions, divided by $3N$, where $N$ is the number of atoms of the atomistic region.

\subsubsection{Results}

\begin{figure}
  \begin{center}
    \includegraphics[scale=0.8]{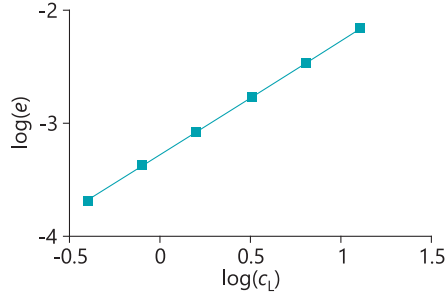}
  \end{center}
  \caption{The dependence of the multiscale modelling error on $c_\mathrm{L}$ for the 1D domain wall propagation example.}
  \label{fig:res_1Derr}
\end{figure}

In figure \ref{fig:res_1Derr}, the dependence of the multiscale modelling error on $c_\mathrm{L}$ is shown. This is the error due to numerical multiscale treatment of the demagnetisation field, compared to the reference case, where the analytical expression for the demagnetisation field is used. The error is proportional to $c_\mathrm{L}$.

\subsection{HMM examples}

\subsubsection{A chain of magnetic particles}

\begin{figure}
  \begin{center}
    \includegraphics[width=0.45\textwidth]{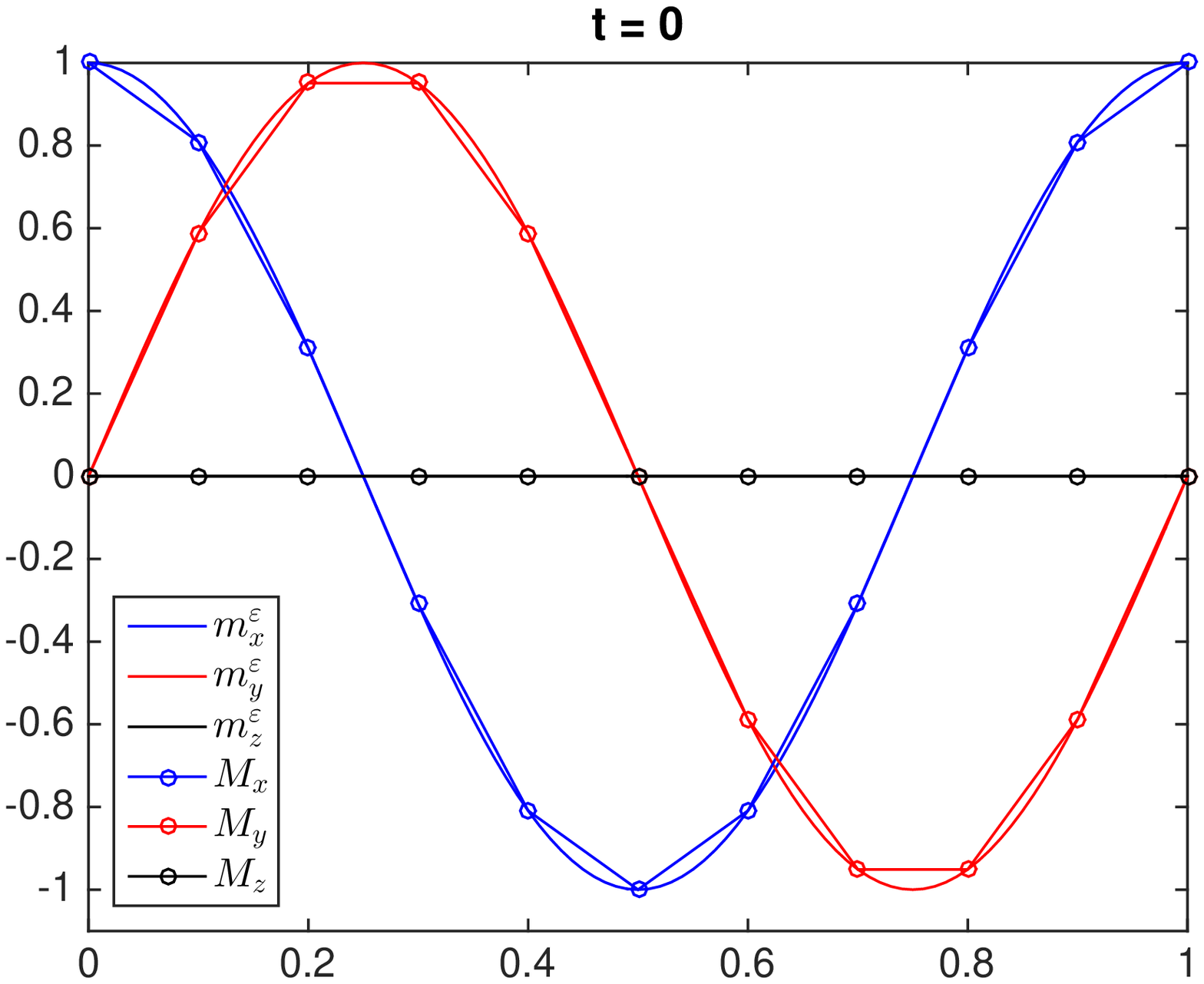}
    \includegraphics[width=0.45\textwidth]{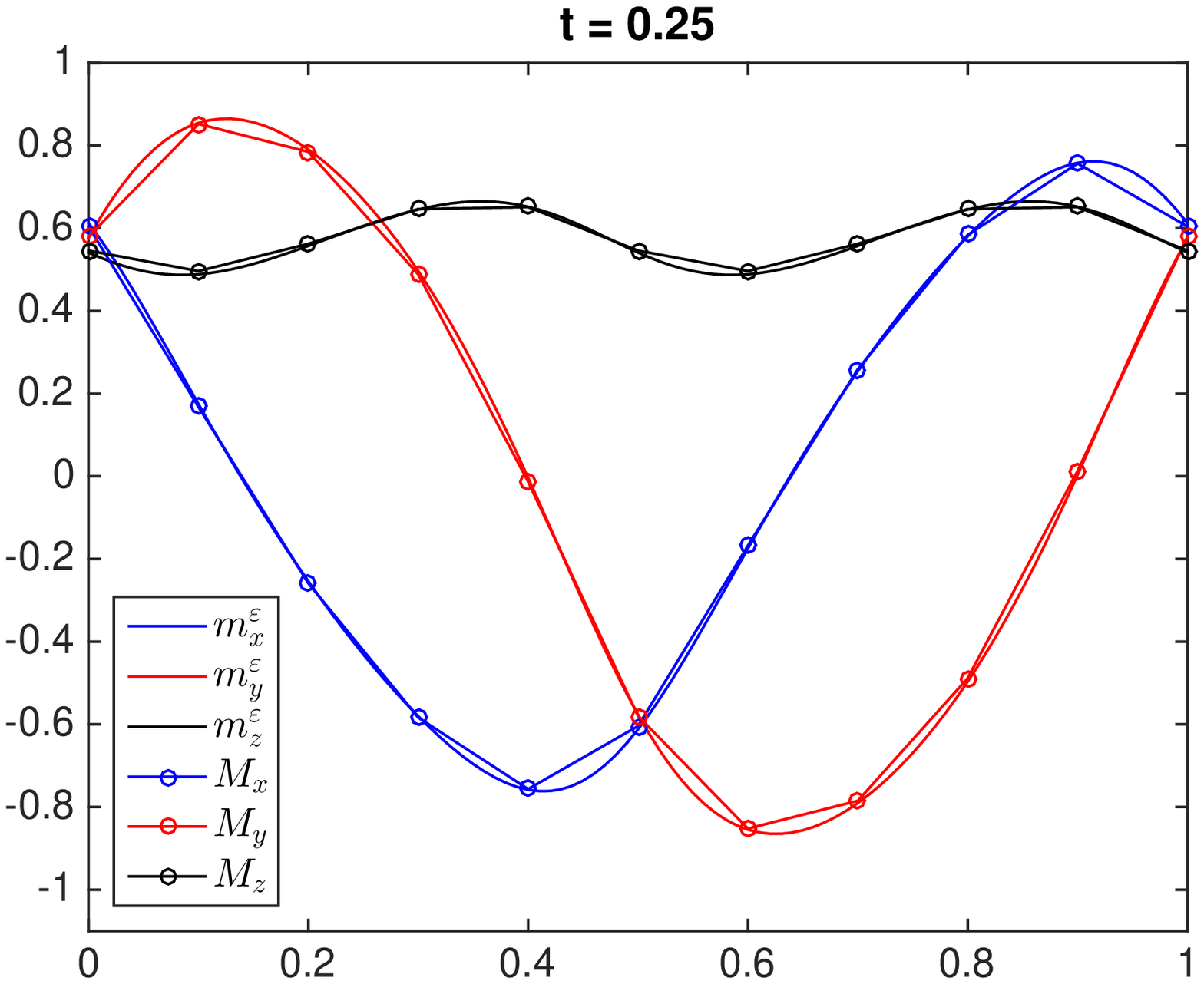}
    \includegraphics[width=0.45\textwidth]{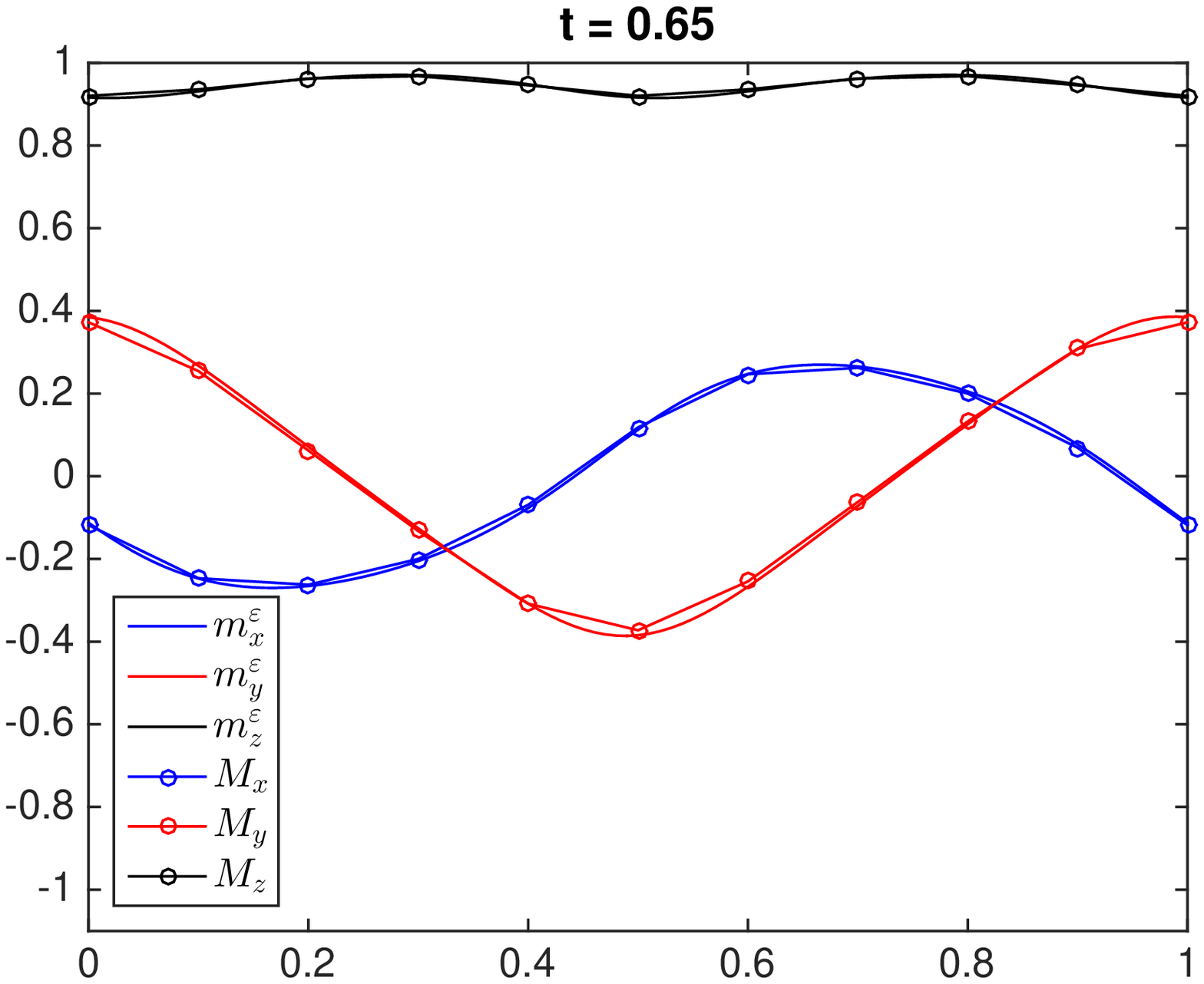}
    \includegraphics[width=0.45\textwidth]{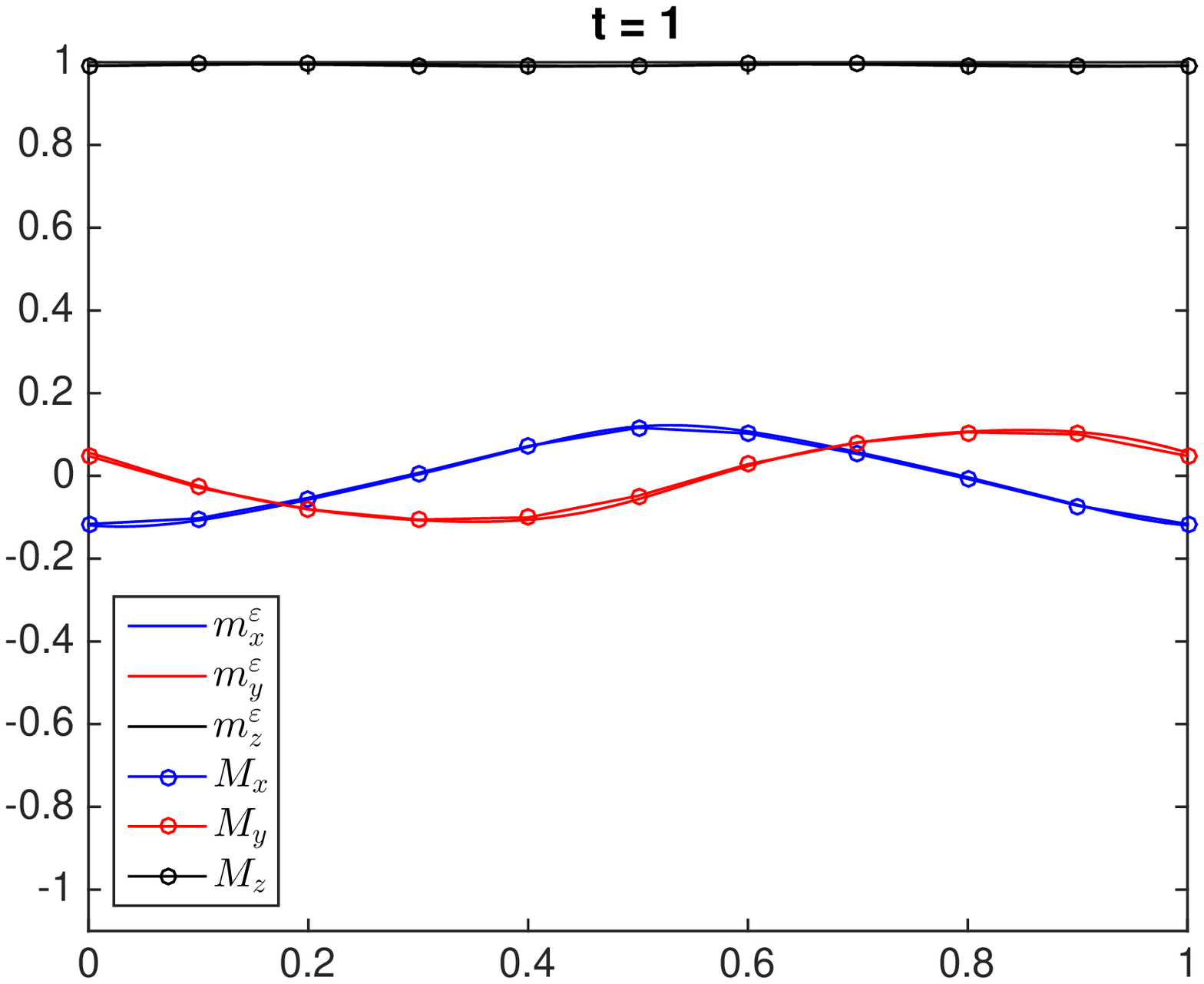}
  \end{center}
  \caption{The HMM solution, using $10$  macroscopic points, is compared to an atomistic simulation using $100$ magnetic moments. A good match between the two solutions are observed.}
  \label{Fig:ZeroTemHMM}
\end{figure}

Here, the HMM algorithm at zero temperature is applied to a chain of magnetic particles. It is assumed that $N= 101$ magnetic moments are located on a one-dimensional lattice consisting of points $ \{ i a \}_{i=0}^{100}$, where $a =0.01$ is the atomic distance and the initial configuration is given by
\begin{equation*}
  \vectorn{m}_{i}(t)|_{t=0} = \cos(2 \pi i a) \vectorn{e}_x + \sin(2 \pi i a) \vectorn{e}_y , \quad i=0,1,\ldots,100 .
\end{equation*}
Moreover, the high frequency external field
\begin{equation*}
  \vectorn{H}_{\mathrm{e}}^{\e}(t) = f^{\e}(t) \vectorn{e}_{z} ,
\end{equation*}
with $f^{\e}(t) = 1 + \cos(0.43 t ) + \cos^2(2\pi t / \e)$ and $\e = 0.01$ is used in the simulations. Note that all the atomic particles are under the influence of the same external field and hence the spatial dependency is omitted. The continuum demagnetisation field $\vectorn{H}_{\mathrm{c}}$ in the microscopic model \eqref{MicroProblemChain_eqn}, is given exactly by 
\begin{equation*}
  \vectorn{H}_{\mathrm{c}} = -M_{x} \vectorn{e}_{x} 
\end{equation*}
in this one-dimensional setting, cf. the derivations in Section \ref{Sec_OneDimAnal}. The parameters $\beta_{\mathrm{L}} = \alpha_{\mathrm{L}} = \mu = 1$, and the nearest neighbour exchange interactions with $J=1$, see Section \ref{Sec:HMM_Zero_Temprature}, are chosen for the simulations. Moreover, the anisotropy is assumed to be zero. The problem is discretised by a midpoint rule on the macroscopic and microscopic scales with the temporal step sizes $\Delta t = 0.01$ on the macro-scale and $\delta t = \e/10$ on the micro-scales. Moreover, the temporal microscopic box is $\tau = 5 \e$. On the spatial dimension, $10$ macroscopic points are used to describe the magnetisation dynamics whereas the atomistic chain consists of a total of $101$ particles. Figure \ref{Fig:ZeroTemHMM} demonstrates the evolution of the macroscopic dynamics, and compares it to the solution of the full atomistic model. In the atomistic simulation it is assumed that $\vectorn{H}_{\mathrm{a},i} = -m_{x}\vectorn{e}_x$, which is justified by the convergence of the atomistic long-range field to the continuum counterpart as $a \to 0$, see Section \ref{subsec:ApproximationError}. As time increases, the magnetisation vectors are pointing to the $\vectorn{e}_{z}$ direction, i.e. the direction of the given external field, and it is shown in the picture that the correct macroscopic dynamics are captured even though the macroscopic discretisation parameters under-resolves the scales of atomistic variations both in time and space. Note that the presence of the long-range field  has a clear effect on the dynamics of the magnetisation, cf.\ the numerical results in \citep{Arjmand_etal_2016} in the absence of the long-range field. 

\subsubsection{Modelling error}

\begin{figure}
  \begin{center}
    \includegraphics[width=0.55\textwidth]{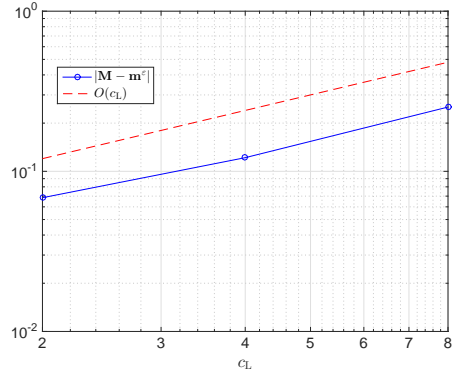}
  \end{center}
  \caption{The error between the HMM solution and the full atomistic simulation is depicted for an increasing values for the long-range field. A first order convergence, in terms of $c_\mathrm{L}$, is observed in the simulation.}
   \label{Fig:LongRangeError}
\end{figure}

The error, which arises from having the long-range field in the modelling is studied. The error between the HMM solution and the atomistic solution is recorded with varying values for the coefficient $c_\mathrm{L}$. The very same numerical parameters as in the previous example are chosen in the simulation. It is observed that the decay rate of the error is $O(c_\mathrm{L})$, see Figure \ref{Fig:LongRangeError}.

\subsubsection{A chain of magnetic particles at nonzero temperature}

\begin{figure}
  \begin{center}
    \includegraphics[width=0.55\textwidth]{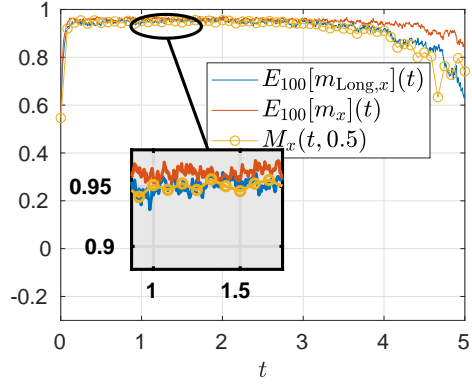}
  \end{center}
  \caption{The $x$ component, $M_{x}(t,X_{I})$, of the HMM solution, at the point $X_{I} = 0.5$ is compared to the full atomistic solution. Here $m_{\mathrm{Long,x}}$ stands for the full atomistic solution, where the long-range is included in the modelling, whereas $m_{x}$ is the atomistic solution without the presence of the long-range interactions.}
  \label{Fig:HMMTemp1}
\end{figure}

\begin{figure}
  \begin{center}
    \includegraphics[width=0.55\textwidth]{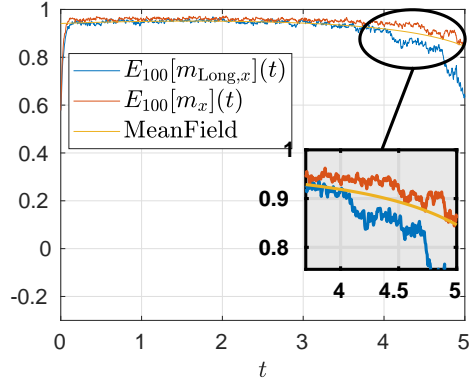}
  \end{center}
  \caption{The atomistic solution with and without the long-range interaction is plotted against the solution of the mean-field approach. The classical mean-field approach breaks down when the effect of the exchange coefficient becomes dominant.}
  \label{Fig:HMMTemp2}
\end{figure}

In this section, the aim is to show an example, where the existence of the long-range interaction has an effect over the macroscopic magnetisation length at elevated temperatures. For a numerical study, $N=101$ atomistic particles located on a one-dimensional lattice consisting of points $ \{ i a \}_{i=0}^{100}$, where $a =0.01$ is the atomic distance, are considered. The initial configuration is uniform and given by 
\begin{equation*}
  \vectorn{m}_{i}(t)|_{t=0} = \dfrac{1}{\sqrt{3}}\left(\vectorn{e}_{x} + \vectorn{e}_{y} + \vectorn{e}_{z} \right), 
    \quad i=0,1,\ldots,100,
\end{equation*}
equipped with periodic boundary conditions. The anisotropy is assumed to be zero, and the external field is pointing in the $x$ direction, i.e. 
\begin{equation*}
  \vectorn{H}^{\e}_{\mathrm{e}} = \left( 1 + \cos(0.43 t) + \sin(0.73 t) + \cos^2 (2 \pi t/\e) \right) \vectorn{e}_x .
\end{equation*}
The atomistic particles are subjected to a thermal noise with a standard deviation of $D = 0.01$, and the parameters $\beta_{\mathrm{L}} = \mu = 1$, $\alpha_{\mathrm{L}} = 10$, with the nearest neighbour interactions for the exchange coefficient, similar to previous examples, are chosen for the simulation. In Figure \ref{Fig:HMMTemp1}, a fully atomistic simulation (with $100$ realisations) with and without the long-range field is compared against the HMM solution up to a final time $T = 5$. It is shown that the long-range field has an effect on the length of the magnetisation and that the HMM accurately captures this long-range effect. In Figure \ref{Fig:HMMTemp2}, the behaviour of the classical mean-field approach is shown. The mean-field approach is a well-known approach to predict the length of the ensemble averages, $|\langle \vectorn{m} \rangle|$, based on the closure argument $E[\vectorn{m}_i \cdot \vectorn{m}_j]  = E[\vectorn{m}_i ] \cdot  E[\vectorn{m}_j]$, and the restrictive assumption that $E[\vectorn{m}_i] = E[\vectorn{m}_j]$ for all $i,j$. The mean-field formula, for this specific setup, reads as 
\begin{multline*}
  |\langle \vectorn{m} \rangle| = |\langle m_{x} \rangle| = \coth\left( \frac{1}{k_{\mathrm{B}} T} \left(
    \mu H^{\e}_{\mathrm{e},x} + 2 J |\langle m_{x} \rangle| - \mu |\langle m_{x} \rangle| \right)\right) \\
    -\dfrac{k_{\mathrm{B}} T}{\mu H^{\e}_{\mathrm{e},x} + 2 J |\langle m_{x} \rangle| - \mu |\langle m_{x} \rangle|} ,
\end{multline*}
see e.g. \cite{APK_Temp_2017,Aharoni1996} for the derivations of the mean-field formulas, which can be adapted to the present setting in an obvious way. It is observed that unlike HMM, the mean-field approach deviates from the true atomistic simulation in the presence of the long-range interaction. This is due to the fact that the mean-field approach  suffers from the mentioned restrictive closure arguments, which do not hold in general since the atomic moments are correlated through the short-range exchange and the long-range dipole-dipole interactions. 

\section{Conclusions}

This paper demonstrates a way of including the long-range dipole-dipole interactions between the atomistic spin magnetic moments into the existing atomistic-continuum coupling methods based on the partitioned-domain and the upscaling strategies. This is achieved by modelling the demagnetising field exclusively at the continuum level and coupling the continuum demagnetising field to the atomistic solution. This approximation relies on the atomistic expression for the magnetisation field converging to the continuum expression when the interatomic spacing approaches zero. It has been demonstrated that in both partitioned-domain and upscaling strategies, the modelling error is $O(c_\mathrm{L})$, where $c_\mathrm{L}$ is the coefficient defining the magnitude of the demagnetising field. Moreover, the present article includes numerical results addressing the convergence of the atomistic long-range field to the continuum field and geometric errors involved in the atomistic simulations of the demagnetisation field.

Within the framework of partitioned-domain methods, it has been discussed how to account for the long-range interactions in the energy-based methods and force-based methods. In both approaches, an auxiliary continuum mesh is constructed that covers the entire computational domain, overlapping with the atomistic region and coinciding the continuum mesh within the continuum region. The equations for the demagnetising field are then solved on this auxiliary mesh. Within the atomistic region, the atomistic demagnetising field is taken to be equal to the continuum demagnetising field.

The computational examples of this paper attempted to highlight cases when the proposed multiscale approaches excel in terms of efficiency. The effect of the void-affected kinetics of the domain wall was modelled using the partitioned-domain approach. Only a small region around the void was modelled at the atomistic scale, while the rest of the 2D structure was modelled using the continuum model with a coarser resolution. The domain wall structure itself was the result of the demagnetising field. The partitioned-domain methodology allows resolving fine-scale details of the interaction of the domain wall and the void, while replacing the solution far from the void with a close continuum approximation to increase computational efficiency.

As for the upscaling strategy, the main novelty and advantage with the proposed algorithm is that the long-range atomistic communication between the magnetic moments, located in different microscopic boxes, is avoided. Nevertheless, the macroscopic effect of the long-range interactions is captured accurately in a multiscale formalism. This leads to a significant computational gain not only in comparison to a naive computation of the dipole-dipole interactions (which scales quadratically with respect to the number of particles) but also when compared to more efficient multiscale methods such as the fast multipole method \cite{Jourdan2008,Greengard_Rokhlin_97}, which is a linear scaling algorithm. The fact that the long-range field is included in the microscopic model, using the continuum field $\vectorn{H}_{\mathrm{c}}$ allows for obtaining sublinear scaling computational costs with respect to the atomistic degrees of freedom. The accuracy of the method is also demonstrated using an example of a chain of magnetic particles as well as examples at elevated temperatures. 

To sum up, an efficient multiscale modelling of the demagnetisation field for two well-established multiscale formalisms, the partitioned domain and the upscaling approaches, has been proposed. The ideas presented in this paper allow for a complete multiscale simulation of ferromagnetic materials, as they can be used when designing efficient modelling strategies taking into account both short- and long-range interactions between the spin magnetic moments and a finite temperature. 

\bibliographystyle{abbrvnat}
\bibliography{refs}

\end{document}